\documentclass[cameraready]{Interspeech}
\usepackage{times}
\usepackage{fontawesome5}
\usepackage{latexsym}
\usepackage{amsthm}   % for theorem-like environments
\usepackage{amsmath}  % for math formatting
\usepackage{amssymb}  % for symbols
\usepackage{algorithmicx} %
\usepackage{algpseudocode} % 
\usepackage[table]{xcolor} %
\usepackage{graphicx}
\usepackage{microtype}
\usepackage{algorithm}
\usepackage{booktabs}
\usepackage{multirow,tabularx,multicol}
\definecolor{aliceblue}{rgb}{0.94, 0.97, 1.0}
\definecolor{lightorange}{RGB}{240, 195, 150}
\definecolor{lightgreen}{RGB}{150, 240, 195}
\definecolor{lightblue}{RGB}{150, 195, 240}
\definecolor{lightred}{RGB}{240,170,170}

\usepackage{subcaption}
\usepackage{soul} % highlight text

\usepackage{pifont}
\usepackage{svg}

\theoremstyle{plain} % bold title, italic body
\newtheorem{theorem}{Theorem}[section]

\theoremstyle{definition} % bold title, upright body
\newtheorem{definition}[theorem]{Definition}

\theoremstyle{remark} % italic title, upright body

\title{DiFlow-TTS: Compact and Low-Latency Zero-Shot Text-to-Speech with Discrete Flow Matching}

\author[affiliation={1}]{Ngoc-Son}{Nguyen}
\author[affiliation={1}]{Thanh V. T.}{Tran}
\author[affiliation={1}]{Hieu-Nghia}{Huynh-Nguyen}
\author[affiliation={2}]{Truong-Son}{Hy}
\author[affiliation={1}, correspondingauthor]{Van}{Nguyen}
\address{
    $^1$ FPT Software AI Center, Vietnam \\
    $^2$ University of Alabama at Birmingham, USA
}

\email{\{sonnn45, thanhtvt1, vannth19\}@fpt.com, 
huynhnguyenhieunghia1999@gmail.com, thy@uab.edu}
\Code{https://github.com/Fsoft-AIC/DiFlowTTS}


\usepackage{comment}

\begin{document}
\maketitle

\begin{abstract}

Zero-shot text-to-speech (TTS) has made significant progress in replicating unseen voices, yet balancing generation quality and inference efficiency remains challenging. Autoregressive models suffer from high latency, while diffusion-based approaches are constrained by training-time configurations. Moreover, most flow-based methods operate in continuous space, which introduces optimization challenges because continuous token spaces are inherently more complex than discrete ones. To address these limitations, we propose \textbf{DiFlow-TTS}, a novel zero-shot TTS framework based on discrete flow matching. The model consists of a deterministic Phoneme-Content Mapper for linguistic modeling and a Factorized Discrete Flow Denoiser that simultaneously generates prosody and acoustic token streams. Experimental results demonstrate the effectiveness of our approach across multiple evaluation metrics.
\end{abstract}
\section{Introduction}
Zero-shot text-to-speech (TTS) has made remarkable progress in recent years, with the goal of generating high-quality speech that faithfully replicates the voice of previously unseen speakers from only a few seconds of reference audio. Neural audio codecs \cite{encodec, zeghidour2021soundstream} have enabled high-fidelity audio compression by efficiently encoding raw waveforms into discrete tokens while preserving quality through the residual vector quantization (RVQ) method.  Recent studies have explored autoregressive (AR) modeled by applying language modeling techniques to these discrete tokens  \cite{voicecraft, melle, valle, spark-tts}, demonstrating high-quality speech generation. Although these methods are widely validated for discrete tokens, they inherently suffer from high inference latency due to their step-by-step sequential generation.

% To overcome these limitations, non-autoregressive (NAR) models have been developed to enable faster generation through parallel decoding. Among these, diffusion-based \cite{kang23_interspeech, ns2, ns3, lee2025dittotts} and flow-based \cite{kim2023pflow, le2023voicebox, matcha-tts, e2tts, f5tts} models in continuous settings have emerged as effective generative frameworks for TTS, striking a balance between synthesis quality and inference efficiency. Nevertheless, their performance still degrades when applied to discrete sequential data compared to AR models.

To address this issue, recent efforts to adapt these discrete codec tokens to non-autoregressive generation have sparked growing interest in applying diffusion models in both discrete settings \cite{diffsound, dctts, instructtts, ns3} and continuous settings \cite{ns2, lee2025dittotts}. However, diffusion-based methods tightly couple the training and sampling processes, restricting sampling configurations to those established during training \cite{qin2025defog}. Modifying components such as noise schedules or rate matrices requires retraining for each new configuration, resulting in considerable computational overhead. Flow matching, on the other hand, offers a more flexible and efficient alternative. Most existing flow-based models still operate in purely continuous representation spaces and follow a single design paradigm, where flow is defined over continuous representations derived from discrete inputs \cite{ozspeech} or mel-spectrograms \cite{matcha-tts, f5tts}. Although these representations are theoretically richer, they span a highly dimensional and unbounded space, which complicates density estimation and increases the likelihood of out-of-distribution artifacts, unlike discrete tokens that exist within a structured and finite space \cite{zheng2025rethinking}. This limitation motivates our exploration of Discrete Flow Matching (DFM) \cite{dfm} for speech synthesis.

\begin{figure}[t]
\centering
\includegraphics[width=\columnwidth]{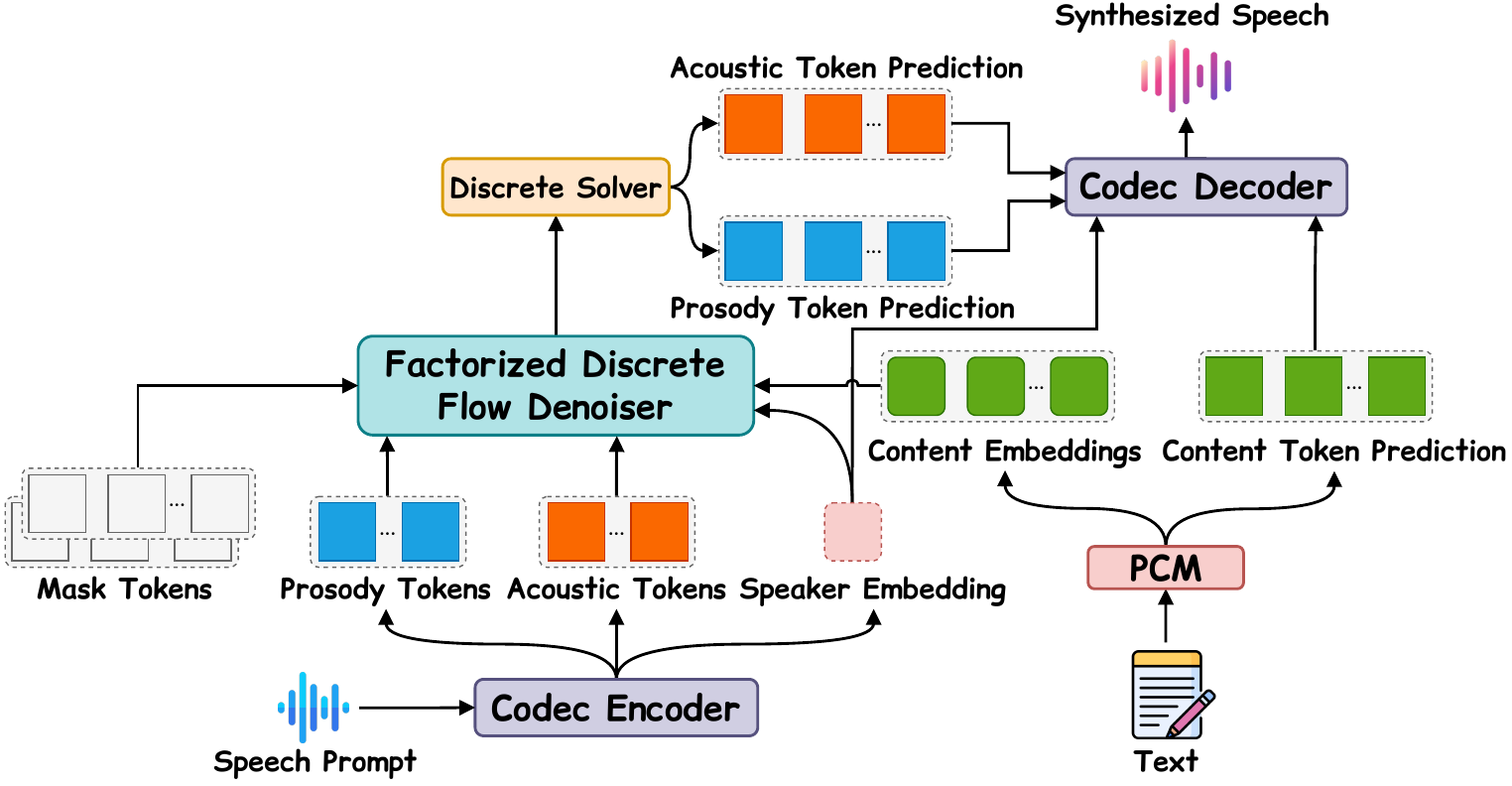}
\caption{\textbf{Overview of DiFlow-TTS}. A Codec Encoder decomposes the speech prompt into a \textbf{speaker embedding}, prosody, and acoustic tokens for zero-shot style transfer, while the Phoneme-Content Mapper converts text into \textbf{content tokens} and content embeddings. Conditioned on the content embeddings and the speaker identity, prosody, and acoustic tokens extracted from the speech prompt, the Factorized Discrete Flow Denoiser simultaneously generates \textbf{prosody and acoustic tokens}. Finally, the generated tokens together with the speaker embedding are fed into the Codec Decoder to reconstruct the waveform.}
\label{fig:overview}
\vspace{-0.2cm}
\end{figure}

In response, we present \textbf{Di}screte \textbf{Flow} Matching for zero-shot \textbf{TTS} (DiFlow-TTS), illustrated in Figure \ref{fig:overview}, a novel zero-shot TTS framework that leverages DFM tailored to discrete settings in speech synthesis. This work aims to demonstrate DFM as a promising research direction for speech synthesis, supported by empirical experiments that reveal both the advantages and limitations of this approach. To address this gap, we take an initial step toward developing a framework that operates directly in the discrete space of factorized codec tokens by employing a pre-trained FACodec~\cite{ns3} as the target discrete representation. This choice is motivated by two key factors: \ding{182} FACodec factorizes speech into prosody, content, and acoustic attributes, enabling flexible modeling of each attribute independently; and \ding{183} it is pre-trained on a large-scale multi-speaker corpus, providing a robust and reliable codec foundation for our target data. Building on this, DiFlow-TTS explicitly models these factorized attributes within a compact and unified framework. Specifically, we design \textbf{P}honeme-\textbf{C}ontent \textbf{M}apper (PCM), which directly maps phoneme sequences to discrete speech tokens that represent the content of the utterance. This module further generates content embeddings that align closely with the semantic structure of the speech. These embeddings, along with auditory attributes extracted from the reference speech prompt, are then used to condition a \textbf{F}actorized \textbf{D}iscrete \textbf{F}low \textbf{D}enoiser (FDFD) module, allowing it to effectively clone the reference's speaking characteristics. Crucially, we design the denoiser with separate prediction heads for the probability velocity of distinct speech aspects, specifically prosody and acoustic details, allowing it to simultaneously learn multiple attribute-specific distributions within a unified flow architecture. In summary, our main contributions are as follows:

\begin{itemize}
    % \item We introduce the first application of discrete flow matching 
    % \item We introduce the first purely discrete flow matching framework for zero-shot TTS, harnessing in-context learning on factorized codec tokens to synthesize high-fidelity speech in a fully NAR manner.
    \item We introduce DiFlow-TTS, a novel zero-shot TTS framework that learns probability flows directly in the discrete space of factorized codec tokens, establishing an initial baseline for applying DFM to speech generation.
    %\item We propose a Factorized Discrete Flow Denoiser that incorporates a factorized flow-prediction mechanism with dedicated heads for prosody and acoustic details, enabling explicit learning of aspect-specific distributions within a compact architecture, without the need for multiple generators.
    % \item We present Phoneme-Content Mapper, which aligns phoneme sequences to discrete content tokens, providing precise semantic grounding that guides the generation of prosody and acoustic attributes.
    \item Unlike prior DFM applications that operate on homogeneous discrete sequences, we propose a reformulated DFM framework over a structured, factorized representation by introducing the Factorized Discrete Flow Denoiser. This design explicitly models individual speech attributes through a flow-prediction mechanism with dedicated heads for prosody and acoustic details. To our knowledge, this is the first work to decompose probability velocity fields across multiple discrete subspaces within a single discrete flow process.
    %\item We demonstrate that DiFlow-TTS outperforms baselines in naturalness, content accuracy, and prosody preservation while maintaining a compact model size, up to 11.7 times smaller, and achieving low-latency inference, up to 34 times faster than baselines, making it suitable for resource-constrained, latency-sensitive systems.
    \item We show that DiFlow-TTS achieves promising results compared to baseline models in terms of naturalness, content accuracy, and prosody preservation, while retaining a compact architecture that is up to $11.7\times$ smaller and delivering low-latency inference with up to $34\times$ speedup over baselines, making it a promising candidate for deployment in resource-constrained and latency-critical environments.
\end{itemize}
\section{Related Work}
A growing trend in speech synthesis focuses on converting raw waveforms into discrete token representations using vector-quantized variational autoencoders (VQ-VAE), which was first introduced by \cite{VQ-VAE} in the field of computer vision and later adapted to speech processing \cite{vq-wav2vec,hubert}. These tokenized representations have demonstrated greater naturalness and robustness compared to conventional mel-spectrogram-based approaches. To effectively model sequences of discrete speech tokens, recent efforts have adapted large language models (LLMs) from the natural language processing domain \cite{voicecraft, melle, valle, spark-tts}. A notable example is VALL-E \cite{valle}, which leverages a pre-trained neural codec to encode speech into discrete codec tokens and reformulates zero-shot TTS as a conditional codec language modeling task. During inference, it performs autoregressive continuation from the acoustic tokens of a speech prompt, enabling high-fidelity speaker-consistent voice synthesis.

Although AR models achieve impressive quality, they are inherently limited by slow inference speeds. This limitation has prompted a shift toward NAR paradigms \cite{ns2, ns3, unicasts, lee2025dittotts, jia2025ditar}. For example, NaturalSpeech 2 \cite{ns2} uses diffusion \cite{NEURIPS2020_4c5bcfec, song2021scorebased} to generate discrete acoustic tokens as continuous features. Its successor, NaturalSpeech 3 \cite{ns3}, further factorizes speech into subspaces of content, prosody, and acoustic details, employing multiple diffusion models to independently capture various acoustic characteristics. In parallel, flow matching \cite{cfm, rectified-flow} has gained attention as a promising generative technique, producing strong results in various domains. Yet, most existing speech-related flow matching applications operate in continuous spaces, using either mel-spectrogram representations or continuous representations derived from discrete tokens, and predict flows over these representations rather than directly over discrete representations. \cite{matcha-tts, guan2024reflow, yao2025stablevc, zuo2025enhancing, ozspeech}.

An emerging line of research seeks to extend iterative refinement techniques to discrete spaces by modeling generation dynamics with Markov chains. Discrete-space generative models have already proven effective in domains such as natural language \cite{pmlr-v235-lou24a, shi2024simplified}, proteins \cite{CampbellYBRJ24, yi2025allatom}, vision \cite{chang2022maskgit, shi2024simplified, fuest2025maskflow}, code \cite{dfm}, and graphs \cite{qin2025defog}. However, the application of DFM \cite{dfm} to speech modeling, such as zero-shot TTS, remains largely unexplored. In this work, we propose a zero-shot TTS system with DFM, aiming to leverage the advantages of discrete modeling while preserving high-quality speech generation.
\section{Methodology}\label{sec:method}
\begin{figure*}[t]
\centering
\includegraphics[width=\textwidth]{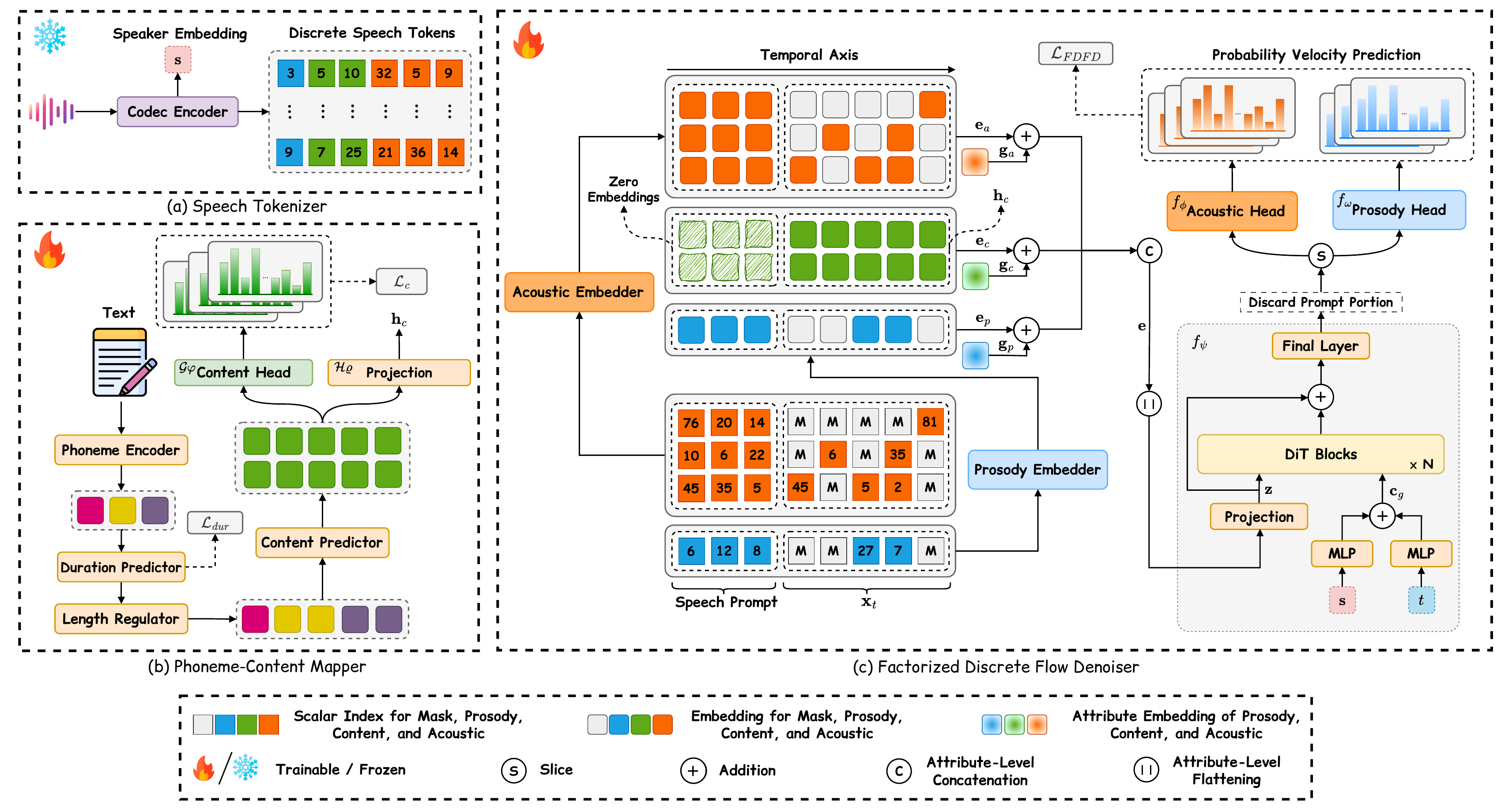} % Reduce the figure size so that it is slightly narrower than the column.
\caption{\textbf{Detailed architecture of DiFlow-TTS.} We formulate zero-shot TTS as token prediction over a factorized codec token space. Speech is tokenized by the \textbf{(a)} \textit{Speech Tokenizer} into content, prosody, and acoustic tokens along with a speaker embedding. Built on these tokens, we design a framework comprising two main modules: \textbf{(b)} \textit{Phoneme-Content Mapper}, which maps input phonemes to discrete content tokens and generates corresponding content embeddings; and \textbf{(c)} \textit{Factorized Discrete Flow Denoiser}, which performs discrete flow matching conditioned on the speaker embedding, the discrete prosody and acoustic tokens derived from the speech prompt, and the content embeddings.}
\label{fig:components}
\end{figure*}

Figure \ref{fig:components} illustrates the details of the DiFlow-TTS framework, which consists of two core modules: \textbf{(b)} \textit{Phoneme-Content Mapper} and \textbf{(c)} \textit{Factorized Discrete Flow Denoiser}, described in detail in Sections~\ref{sec:pcm} and~\ref{sec:fdfd}, respectively. We first employ a pre-trained \textbf{(a)} \textit{Speech Tokenizer}, described in Section~\ref{sec:tokenizer}, to extract discrete token sequences as our target representations, which are subsequently modeled by these two modules.

% \begin{figure}[t]
% \centering
% \includegraphics[width=\columnwidth]{figures/content_predictor.pdf}
% \caption{The detailed architecture of the Content Predictor.}
% \label{fig:content_predictor}
% \end{figure}

\begin{figure}[t]
\centering
\includegraphics[width=0.6\columnwidth]{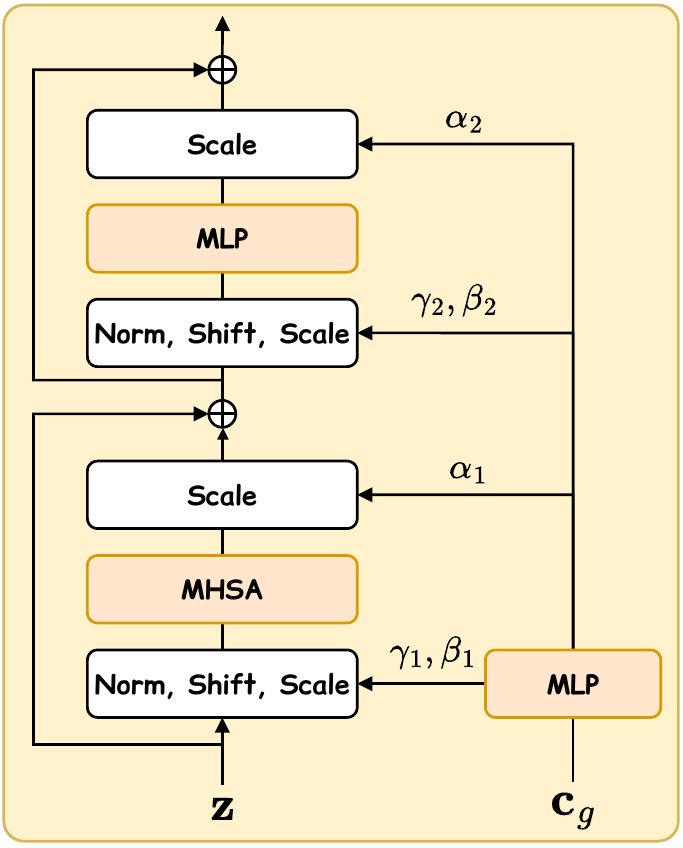}
\caption{The detailed architecture of the
DiT block.}
\label{fig:dit}
\end{figure}

% \begin{figure*}[t]
% \centering
% \begin{minipage}{0.60\textwidth}
%     \centering
%     \includegraphics[width=\linewidth]{figures/content_predictor.pdf}
%     \captionof{figure}{The detailed architecture of the Content Predictor.}
%     \label{fig:content_predictor}
% \end{minipage}
% \hfill
% \begin{minipage}{0.30\textwidth}
%     \centering
%     \includegraphics[width=\linewidth]{figures/dit}
%     \captionof{figure}{DiT block architecture.}
%     \label{fig:dit}
% \end{minipage}

% \hfill
% \vspace{-0.35cm}
% \end{figure*}

\subsection{Notation}
Let a sequence $x$ be an array of $L$ tokens $(x_1, x_2, \dots, x_L)$ drawn from a discrete vocabulary of size $v$, i.e., $x \in [v]^L$ with $[v] = \{1, \dots, v\}$. To represent point mass distributions over these sequences, we use the delta function $\delta_y(x) = \prod_{i=1}^L \delta_{y_i}(x_i)$, where $y \in [v]^L$; $\delta_{y_i}(x_i) = 1$ if $x_i = y_i$, and $0$ otherwise. Building on this, we further define the extended space $[v]^{nL}$ to represent the concatenation of $n$ sequences, each of length $L$.

\subsection{Input Data Extraction}\label{sec:tokenizer}
% The {Speech Tokenizer} module (Figure \ref{fig:components}a) converts a raw speech waveform into discrete token sequences. For this process, we employ FACodec \cite{ns3}, which factorizes the original speech signal $\mathbf{r}$ into disentangled token sequences representing prosody, content, and acoustic details and extracts the speaker identity:
We employ FACodec \cite{ns3}, comprising a Codec Encoder and a Codec Decoder, as our \textit{Speech Tokenizer} (Figure \ref{fig:components}a) that converts raw speech waveforms into discrete token sequences serving as target representations in our framework. Specifically, the Codec Encoder converts the original speech signal $\mathbf{r}$ into disentangled token sequences representing prosody, content, and acoustic details and extracts the speaker embedding:
\begin{equation}
\mathbf{r}^p, \mathbf{r}^c, \mathbf{r}^a, \mathbf{s} = \text{CodecEncoder}(\mathbf{r}),
\label{eq:tokenizer}
\end{equation}
where $\mathbf{r}^p \in [v]^{L}$, $\mathbf{r}^c \in [v]^{2L}$, and $\mathbf{r}^a \in [v]^{3L}$ denote the token sequences for prosody, content, and acoustic details, respectively, and $\mathbf{s} \in \mathbb{R}^{D_{\text{spk}}}$ is the speaker embedding. Here, $L$ is the token sequence length, and prosody, content, and acoustic details use one, two, and three Residual Vector Quantization (RVQ) codebooks, respectively, each with a vocabulary size of $v = 1024$.

We design a model that generates these tokens ($\hat{\mathbf{r}}^p \in [v]^{L}, \hat{\mathbf{r}}^c \in [v]^{2L}, \hat{\mathbf{r}}^a \in [v]^{3L}$) and synthesizes the speech waveform $\hat{\mathbf{r}}$ by feeding the generated tokens together with the speaker identity $\mathbf{s}$  into the Codec Decoder:
\begin{equation}
\hat{\mathbf{r}} = \text{CodecDecoder}(\hat{\mathbf{r}}^p, \hat{\mathbf{r}}^c, \hat{\mathbf{r}}^a, \mathbf{s}).
\end{equation}

\subsection{Discrete Flow Matching for Factorized Codec Tokens}
% \subsubsection{Notation}
% Let a sequence $x$ be an array of $L$ tokens $(x^1, x^2, \dots, x^L)$ drawn from a discrete vocabulary of size $v$, i.e., $x \in \mathcal{D} = [v]^L$ with $[v] = \{1, \dots, v\}$. We further define the extended space $\mathcal{D}' = [v]^{nL}$ for the concatenation of $n$ such sequences. To represent the point mass distributions over these sequences, we use the delta function $\delta_y(x) = \prod_{i=1}^L \delta_{y^i}(x^i)$, where $y \in \mathcal{D}, \delta_{y^i}(x^i) = 1$ if $x^i = y^i$, and $0$ otherwise.

We adopt discrete flow matching as the generative backbone for modeling prosody and acoustic codec tokens. The goal is to transport source samples $\mathbf{x}_0 \sim p$ to target samples $\mathbf{x}_1 \sim q$. Following \cite{dfm}, we instantiate the source distribution $p$ to assign all probability mass to sequences in which every token is the mask token \verb|[MASK]|, that is, $p(x) = \delta_{\verb|[MASK]|}(x)$. This implies that the source distribution places all probability mass in the sequence where every token is the mask token \verb|[MASK]|. For the target distribution $q$, unlike prior DFM applications that operate on homogeneous discrete sequences, we reformulate the DFM framework by factorizing $\mathbf{x}_1$ into two structured components that are learned jointly. This structured formulation enables the construction of a probability velocity field over a composite target space consisting of these two components. Accordingly, we define the target distribution $q$ as follows:
\begin{definition}
Let $\mathbf{x}_1^{p} \sim q_p$ and $\mathbf{x}_1^{a} \sim q_a$ denote the random variables corresponding to the prosody and acoustic details sequences, respectively. These sequences are in spaces $[v]^{L}$ and $[v]^{3L}$. The full target sequence is then defined as $\mathbf{x}_1 = [\mathbf{x}_1^{p} ; \mathbf{x}_1^{a}] \in [v]^{(1+3)L}$, where $[;]$ denotes the concatenation of the sequence. Assuming the independence between the two components, the joint target distribution is factorized as $q(x) = q_p(x^p) \cdot q_a(x^a)$, where $x = [x^p ; x^a]$.
\end{definition}

During training, we employ a scheduler $\kappa_t \in [0,1]$, a monotonically increasing function with boundary conditions $\kappa_0 = 0$ and $\kappa_1 = 1$, where $t \in [0,1]$ denotes continuous time. This scheduler controls the interpolation, gradually shifting the distribution from source to target as $\kappa_t$ increases. We follow the deterministic convex-interpolant assumption, under which the marginal path distribution is given by $p_t(\mathbf{x}^i | \mathbf{x}_t) = \delta_{\mathbf{x}_t}(\mathbf{x}^i)$. Following \cite{dfm}, we then construct a conditional probability path, referred to as the \textit{mixture path}, which linearly interpolates between the source and target distributions:
$
    p_t(\mathbf{x}^i | \mathbf{x}_0, \mathbf{x}_1) = (1 - \kappa_t)\delta_{\mathbf{x}_0}(\mathbf{x}^i) + \kappa_t\delta_{\mathbf{x}_1}(\mathbf{x}^i).
$
This formulation leads to a \textit{conditional probability path}, which is governed by the probability velocity $\mathbf{u}_t$ defined as:
\begin{equation}
\label{eq:u_t}
\mathbf{u}^i_t(\mathbf{x}^i,\mathbf{x}_t) = \frac{\dot{\kappa}_t}{1-\kappa_t}\left[p_{1|t}(\mathbf{x}^i|\mathbf{x}_t, \mu; \theta) - \delta_{\mathbf{x}_t}(\mathbf{x}^i)\right],
\end{equation}
where $\dot{\kappa}_t$ is the time derivative of the scheduler $\kappa_t$, $\theta$ denotes learnable parameters of a probability denoiser, $p_{1|t}(\cdot|\mathbf{x}_t, \mu;\theta)$ is the posterior distribution $\mathbf{x}_1$ given a partially corrupted sequence $\mathbf{x}_t$ and $\mu$ representing a set of multimodal conditioning inputs.

\subsection{Phoneme-Content Mapper} \label{sec:pcm}
The PCM module (Figure~\ref{fig:components}b) aligns phoneme sequences derived from the text prompt with discrete content tokens. While its overall structure is inspired by conventional duration-based alignment mechanisms \cite{ren2021fastspeech}, we reformulate the alignment process to operate directly on discrete codec tokens rather than continuous mel-spectrogram frames.

Given a text input, we convert it into a phoneme sequence and extract phoneme embeddings $\mathbf{p} \in \mathbb{R}^{N \times D}$ using a phoneme encoder, where $N$ and $D$ denote the sequence length and hidden dimension, respectively. To align phonemes with discrete speech tokens, a \textit{Duration Predictor}  estimates durations, indicating how many speech tokens correspond to each phoneme. This produces an integer-based alignment that maps each phoneme to a variable-length span in the speech-token sequence. Using these alignments, the \textit{Length Regulator} upsamples phoneme embeddings based on the ground-truth (during training) or predicted (during inference) phoneme durations. The upsampled sequence, whose length is $L$, is then passed to the \textit{Content Predictor}, which consists of 2 Feed-Forward Transformer (FFT) layers. These layers hierarchically extract textual representations at each level, producing hidden states $\mathbf{h} \in \mathbb{R}^{2 \times L \times D}$, which are then processed by two branches: a projection layer $\mathcal{H}{\varrho}(\cdot)$ produces content embeddings, and a content head $\mathcal{G}{\varphi}(\cdot)$ that outputs logits over a vocabulary of size $v$:
\begin{equation}
\begin{split}
\mathbf{h}_c &= \mathcal{H}_{\varrho}(\mathbf{h}) \in \mathbb{R}^{2 \times L \times D}, \\
p(\mathbf{x}^c | \mathbf{h}; \varphi) &= \text{Softmax}(\mathcal{G}_{\varphi}(\mathbf{h})) \in \mathbb{R}^{2 \times L \times v}.
\end{split}
\label{eq:h_c}
\end{equation}
\subsection{Factorized Discrete Flow Denoiser} \label{sec:fdfd}
The FDFD (Figure \ref{fig:components}c) aims to generate the prosody and acoustic sequences of the synthesized speech by leveraging DFM and in-context learning, conditioned on a set of contextual inputs. In the following, we detail the key elements of this module.

\subsubsection{Contextual Modeling}
We now elaborate on the construction of the conditioning context $\mu$ introduced in Eq.~\eqref{eq:u_t} and describe how it is integrated into our framework.

\ding{182} \textbf{Inputs:} Given a reference speech prompt $\mathbf{r}$, we apply the operation in Eq.~\eqref{eq:tokenizer} to extract \textbf{prosody ($p$)} token sequences $\mathbf{r}^p \in [v]^{L_r}$, \textbf{acoustic ($a$)} token sequences $\mathbf{r}^a \in [v]^{3L_r}$, and a speaker embedding $\mathbf{s} \in \mathbb{R}^{D_{\text{spk}}}$, where $L_r$ denotes the temporal length of the reference prompt and $D_{\text{spk}}$ is the hidden dimension of the speaker embedding. The corrupted input at timestep $t$, $\mathbf{x}_t \in [v]^{(1+3)L}$, is split into prosody tokens $\mathbf{x}_t^p \in [v]^{L}$ and acoustic tokens $\mathbf{x}_t^a \in [v]^{3L}$. We then employ specific embedders for prosody and acoustic (implemented as learned embedding tables) to map each discrete token index from the vocabulary $[v]$ to $D$-dimensional hidden representations $\mathbf{e}^{\alpha}_r$ (\textbf{reference}) and $\mathbf{e}^{\alpha}_t$ (\textbf{corrupted}) for $\alpha \in \{p, a\}$.

% We then use prosody and acoustic embedders, denoted $\mathcal{E}_{p}(\cdot)$ and $\mathcal{E}_a(\cdot)$, to convert these sequences into hidden representations:
% \begin{equation*}
% \scriptstyle
% \begin{split}
% \mathbf{e}^p_r &= \mathcal{E}_p(\mathbf{r}^p) \in \mathbb{R}^{m \times L_p \times D},
% \mathbf{e}^p_t = \mathcal{E}_p(\mathbf{x}^p_t) \in \mathbb{R}^{m \times L \times D}, \\
% \mathbf{e}^a_r &= \mathcal{E}_a(\mathbf{r}^a) \in \mathbb{R}^{k \times L_p \times D},
% \mathbf{e}^a_t = \mathcal{E}_a(\mathbf{x}^a_t) \in \mathbb{R}^{k \times L \times D}.
% \end{split}
% \end{equation*}
\ding{183} \textbf{Conditioning:} We concatenate the reference and corrupted embeddings along the temporal dimension for each attribute, where the reference embeddings provide contextual information for the corrupted embeddings. We further introduce the \textbf{content ($c$)} embedding (Eq.~\eqref{eq:h_c}) as the content conditioning signal ($\mathbf{e}^c_t = \mathbf{h}_c$) to provide linguistic guidance for prosody and acoustic modeling, while the reference content representation is replaced with zeros ($\mathbf{e}^c_r = \mathbf{0}$), since content is not transferred from the reference speech during style transfer. The concatenated embeddings for each attribute is given by $\mathbf{e}_\alpha = [\mathbf{e}_r^\alpha;\mathbf{e}_t^\alpha] \in \mathbb{R}^{\beta \times (L_r + L) \times D}$, where $[;]$ denotes the concatenation operator, and $(\alpha, \beta) \in \{(p, 1), (c, 2), (a, 3)\}$.
% \begin{equation*}
% \begin{split}
% \mathbf{e}_p &= \mathbf{e}^p_r \oplus \mathbf{e}^p_t \in \mathbb{R}^{m \times (L_p + L) \times D}, \\
% \mathbf{e}_c &= \mathbf{e}^c_r \oplus \mathbf{e}^c_t \in \mathbb{R}^{n \times (L_p + L) \times D},\\
% \mathbf{e}_a &= \mathbf{e}^a_r \oplus \mathbf{e}^a_t \in \mathbb{R}^{k \times (L_p + L) \times D},
% \end{split}
% \end{equation*} 

\ding{184} \textbf{Constructing unified embedding:} 
We first introduce a learnable attribute-type embedding $\mathbf{g}_\alpha \in \mathbb{R}^{1 \times 1 \times D}$ for each attribute $\alpha \in \{p, c, a\}$, allowing the model to explicitly differentiate among the corresponding attributes.
Each such embedding is added to the embeddings of its corresponding attribute, yielding $\tilde{\mathbf{e}}_\alpha = \mathbf{e}_\alpha + \mathbf{g}_\alpha$. The resulting embeddings are concatenated along the attribute dimension, yielding the final unified embedding
$
    \mathbf{e} = [\tilde{\mathbf{e}}_p ; \tilde{\mathbf{e}}_c ; \tilde{\mathbf{e}}_a] \in \mathbb{R}^{(1+2+3) \times (L_r + L) \times D}
$.

% $
%     \mathbf{e} = \tilde{\mathbf{e}}_p \oplus \tilde{\mathbf{e}}_c \oplus \tilde{\mathbf{e}}_a
% $, where $\mathbf{e} \in \mathbb{R}^{(m+n+k) \times (L_p + L) \times D}$.

% $
%     \mathbf{e} =  \left[(\mathbf{e}_p + \mathbf{g}_p)\oplus (\mathbf{e}_c + \mathbf{g}_c) \oplus (\mathbf{e}_a + \mathbf{g}_a)\right] \in \mathbb{R}^{(m + n + k) \times (L_p + L) \times D}
% $.
\subsubsection{Factorized Flow Prediction}
We first flatten $\mathbf{e}$ along the attribute dimension, yielding a tensor $\in \mathbb{R}^{(L_r + L) \times (1 + 2 + 3)D}$, and then linearly project it to $\mathbf{z} \in \mathbb{R}^{(L_r + L) \times D}$. The resulting representation is processed by a neural network $f_\psi: \mathbb{R}^{(L_r + L) \times D} \to \mathbb{R}^{(L_r + L) \times (1 + 3)D}$ composed of Diffusion Transformer (DiT) blocks \cite{dit}, a long skip connection, and a final transformation layer (as shown in Figure \ref{fig:dit}). To achieve speaker adaptation, we form a global conditioning vector $\mathbf{c}_g \in \mathbb{R}^D$ by summing two $D$-dimensional embeddings: the speaker embedding, obtained by projecting $\mathbf{s}$ through a Multi-Layer Perceptron (MLP), and the timestep embedding $\mathbf{t}$, obtained by mapping the timestep through an MLP. This vector modulates the DiT features via adaptive layer normalization (AdaLN) \cite{dit}. After processing the DiT blocks via AdaLN, we apply a long skip connection by adding $\mathbf{z}$ to the final DiT output, followed by the final transformation layer comprising layer normalization, AdaLN-based modulation conditioned on $\mathbf{c}_g$, and a linear projection to dimension $(1+3)D$. We then discard the reference portion and permute the result to yield the final hidden representation $
\mathbf{h}_{p,a} \in \mathbb{R}^{(1+3) \times L \times D}$.

To effectively enable the model to jointly attend to information from different representation subspaces, we propose a factorized flow prediction mechanism based on multi-head prediction. In this design, FDFD simultaneously models multiple aspects of speech, specifically prosody and acoustic details. Formally, we define two parallel heads: the \textit{prosody head} $f_{\phi}(\cdot)$ and the \textit{acoustic head} $f_{\omega}(\cdot)$, which independently predict probability distributions corresponding to prosody and acoustic attributes. We first slice the representation $\mathbf{h}_{p,a}$ into two parts: \ding{182} the prosody representation $\mathbf{h}_p \in \mathbb{R}^{1 \times L \times D}$ and \ding{183} the acoustic representation $\mathbf{h}_a \in \mathbb{R}^{3 \times L \times D}$. Each part is processed by its respective head, $f_\phi(\cdot)$ and $f_\omega(\cdot)$, producing logits of the shapes $\mathbb{R}^{1 \times L \times v}$ and $\mathbb{R}^{3 \times L \times v}$, respectively. These logits correspond to the categorical distributions predicted over the discrete token vocabulary for each attribute. Finally, we concatenate the two outputs along the attribute dimension to obtain a tensor in $\mathbb{R}^{(1+3) \times L \times v}$ that serves as the estimated posterior distribution over $\mathbf{x}_1$.

% \subsubsection{Overall Factorized Discrete Flow Denoiser} We define the overall model $f_{\theta}$ as a composition of three main components: a \textit{neural network} $f_{\psi}$, a \textit{prosody head} $f_{\phi}$, and an \textit{acoustic head} $f_{\omega}$. Formally, the model can be expressed as $f_{\theta} = (f_{\phi} \oplus f_{\omega}) \circ f_{\psi}$, where $\circ$ denotes composition.

\subsection{Training Objective}
The overall training objective consists of three loss components corresponding to different modules in our framework. First, we optimize the \textit{Duration Predictor} using the Mean Squared Error (MSE) loss on the logarithmic scale, denoted as $\mathcal{L}_{dur}$, which compares the predicted and ground-truth durations. Second, for the \textit{Content Predictor} defined in Eq.~\eqref{eq:h_c}, we use a cross-entropy loss $\mathcal{L}_c$ between the predicted logits and the discrete content tokens obtained from the ground truth. Third, for the \textit{FDFD} module, we learn a probabilistic denoiser $p_{1|t}$ trained to recover masked tokens under varying masking ratios. The objective is to minimize the cross-entropy loss:
\begin{equation*}
    \mathcal{L}_{FDFD} = -\sum_{i \in \mathcal{T}}\mathbb{E}_{t \sim \mathcal{U}[0,1],(\mathbf{x}_0,\mathbf{x_1}),\mathbf{x}_t}\left[\log p_{1|t}(\mathbf{x}_1^i|\mathbf{x}_t, \mu; \theta)\right], 
\end{equation*}
where $\mathcal{T} = [(1+3)L]$, $\mathbf{x}_t \sim p_t(\mathbf{x}|\mathbf{x}_0, \mathbf{x}_1), \mathbf{x}_0 \sim p$, and $\mathbf{x}_1 \sim q$.
Finally, the total loss is defined as:
\begin{equation}
\mathcal{L} = \lambda_{dur} \mathcal{L}_{dur} + \lambda_c \mathcal{L}_c + \lambda_{FDFD} \mathcal{L}_{FDFD},
\label{eq:total_loss}
\end{equation}
where $\lambda_{dur}$, $\lambda_c$, and $\lambda_{FDFD}$ are hyperparameters that control
the relative importance of each loss term.
\section{Experimental Setup}\label{sec:exp_setup}

\begin{table*}
\centering
\caption{Performance on the \textit{LibriSpeech test-clean} dataset using 3-second audio prompts. $[\diamond]$ means reproduced results, $[\dagger]$ and $[\ddagger]$ mean results inferred from official and unofficial checkpoints, respectively. The best and second best are \textbf{bold} and \underline{underlined}, respectively. WER is reported as a value in the range $[0, 1]$. Abbreviation: E (Emilia), GS (GigaSpeech), LT (LibriTTS).}
\resizebox{0.9\textwidth}{!}{
\begin{tabular}{c|l|c|ccccccc}
\toprule[1.25pt]
\multirow{3}{*}{\textbf{Type}} & \multirow{3}{*}{\textbf{Model}} & \multirow{3}{*}{\textbf{Data (hours)}} & \multirow{3}{*}{\textbf{UTMOS} $\uparrow$}& \multirow{3}{*}{\textbf{WER} $\downarrow$} & \multirow{3}{*}{\textbf{SIM-O} $\uparrow$} & \multicolumn{2}{c}{\textbf{F0}} & \multicolumn{2}{c}{\textbf{Energy}} \\
\cmidrule(lr){7-8} \cmidrule(lr){9-10}
 & &  &  &  & & \textbf{Accuracy} $\uparrow$ & \textbf{RMSE} $\downarrow$& \textbf{Accuracy} $\uparrow$& \textbf{RMSE} $\downarrow$ \\
\midrule
- & Ground Truth & - & 4.10 & 0.02 & - & - & - & - & - \\
\midrule
\multirow{2}{*}{(i)} & VoiceCraft $[\dagger]$ \textcolor{gray}{ACL'24} & GS (9K) & 3.55 & 0.18 & 0.51 & 0.78 & 17.22 & 0.44 & \underline{0.010} \\
& VALL-E $[\diamond]$ \textcolor{gray}{TASLPRO'25} & LT (500) & 3.68 & 0.19 & 0.40 & 0.75 & 21.66 & 0.36 & 0.020 \\
\midrule
\multirow{3}{*}{{(ii)}} & NaturalSpeech 2 $[\ddagger]$ \textcolor{gray}{ICLR'24} & LT (585) & 2.38 & \underline{0.09} & 0.31 & 0.80 & 15.62 & 0.25 & 0.020 \\
& F5-TTS $[\diamond]$ \textcolor{gray}{ACL'25} & LT (500) & 3.76 & 0.24 & 0.52 & 0.80 & 13.78 & 0.67 & \underline{0.010} \\
& F5-TTS $[\dagger]$ \textcolor{gray}{ACL'25} & E (100K) & 3.72 & \underline{0.09} & \underline{0.66} & \underline{0.83} & 12.66 & 0.66 & \underline{0.010} \\
& OZSpeech $[\dagger]$ \textcolor{gray}{ACL'25} & LT (500) & 3.15 & \textbf{0.05} & 0.40 & 0.81 & \underline{11.96} & 0.67 & \underline{0.010} \\
\midrule
{(iii)} & MaskGCT $[\dagger]$ \textcolor{gray}{ICLR'25} & E (100K) & \underline{3.83} & \underline{0.09} & \textbf{0.67} & 0.77 & 14.33 & \textbf{0.75} & \textbf{0.007} \\
\midrule
\rowcolor{aliceblue}
{(iv)} & \textbf{DiFlow-TTS (Ours)} & \textbf{LT (470)} & \textbf{3.98} & \textbf{0.05} & 0.45 & \textbf{0.88} & \textbf{7.97} & \underline{0.73} & \textbf{0.007}\\
\bottomrule[1.25pt]
\end{tabular}}
\label{tab:overall-results-3s}
%\vspace{-0.25cm}
\end{table*}
% \begin{table}
% \centering
% \resizebox{0.6\textwidth}{!}{%
% \begin{tabular}{l|ccc}
% \toprule[1.25pt]
% \textbf{Model} & \textbf{Naturalness} $\uparrow$ & \textbf{Intelligibility} $\uparrow$ & \textbf{Similarity} $\uparrow$
% \\
% \midrule
% Ground Truth & 4.42 $\pm$ 0.12 & 4.54 $\pm$ 0.11 & 4.29 $\pm$ 0.14 \\
% \midrule
% % Spark-TTS \citep{spark-tts} & \textbf{4.44} $\pm$ \textbf{0.14} & \textbf{4.53} $\pm$ \textbf{0.11} & \textbf{4.54} $\pm$ \textbf{0.12} \\
% VoiceCraft & 3.94 $\pm$ 0.17 & 4.08 $\pm$ 0.18 & 4.17 $\pm$ 0.15  \\
% VALLE-E & 3.71 $\pm$ 0.17 & 3.96 $\pm$ 0.17 & 3.99 $\pm$ 0.15 \\
% NaturalSpeech 2 & 2.62 $\pm$ 0.20 & 3.25 $\pm$ 0.21 & 2.63 $\pm$ 0.18 \\
% F5-TTS & 3.97 $\pm$ 0.17 & 4.16 $\pm$ 0.14 & 4.07 $\pm$ 0.16 \\
% OZSpeech & 2.80 $\pm$ 0.23 & 3.42 $\pm$ 0.24 & 3.20 $\pm$ 0.22 \\
% MaskGCT & ? $\pm$ ? & ? $\pm$ ? & ? $\pm$ ? \\
% \midrule
% \rowcolor{aliceblue}
% \textbf{DiFlow-TTS} & \textbf{4.18 $\pm$ 0.16} & \textbf{4.41 $\pm$ 0.13} & \textbf{4.42 $\pm$ 0.12} \\
% \bottomrule[1.25pt]
% \end{tabular}
% }
% \caption{MOS evaluation with 3-second audio prompts, including 95$\%$ confidence intervals. The best and second best are \textbf{bolded} and \underline{underlined}, respectively.}
% \label{tab:mos-results}
% %\vspace{-0.5cm}
% \end{table}

\begin{table}
\vspace{-0.2cm}
\centering
\caption{MOS evaluation with 3-second audio prompts, including 95$\%$ confidence intervals. The best and second best are \textbf{bolded} and \underline{underlined}, respectively.}
\resizebox{\columnwidth}{!}{
\begin{tabular}{l|ccc}
\toprule[1.25pt]
\textbf{Model} & \textbf{Naturalness} $\uparrow$ & \textbf{Intelligibility} $\uparrow$ & \textbf{Similarity} $\uparrow$
\\
\midrule
Ground Truth & 4.42 $\pm$ 0.12 & 4.54 $\pm$ 0.11 & 4.29 $\pm$ 0.14 \\
\midrule
% Spark-TTS \citep{spark-tts} & \textbf{4.44} $\pm$ \textbf{0.14} & \textbf{4.53} $\pm$ \textbf{0.11} & \textbf{4.54} $\pm$ \textbf{0.12} \\
VoiceCraft & 3.94 $\pm$ 0.17 & 4.08 $\pm$ 0.18 & \underline{4.17 $\pm$ 0.15}  \\
VALL-E & 3.71 $\pm$ 0.17 & 3.96 $\pm$ 0.17 & 3.99 $\pm$ 0.15 \\
NaturalSpeech 2 & 2.62 $\pm$ 0.20 & 3.25 $\pm$ 0.21 & 2.63 $\pm$ 0.18 \\
F5-TTS & 3.97 $\pm$ 0.17 & \underline{4.16 $\pm$ 0.14} & 4.07 $\pm$ 0.16 \\
OZSpeech & 2.80 $\pm$ 0.23 & 3.42 $\pm$ 0.24 & 3.20 $\pm$ 0.22 \\
MaskGCT & \underline{3.97 $\pm$ 0.16} & 4.14 $\pm$ 0.15 & \underline{4.17 $\pm$ 0.15} \\
\midrule
\rowcolor{aliceblue}
\textbf{DiFlow-TTS (Ours)} & \textbf{4.18 $\pm$ 0.16} & \textbf{4.41 $\pm$ 0.13} & \textbf{4.42 $\pm$ 0.12} \\
\bottomrule[1.25pt]
\end{tabular}
}
\label{tab:mos-results}
\end{table}

\begin{table*}[!httb]
\centering
\caption{Comparison of model size and latency. For \textbf{DiFlow-TTS}, we report practical low-latency configurations (NFE = 4, 16). The 128-NFE setting used in Table~\ref{tab:overall-results-3s} achieves RTF = 0.39 (see Table~\ref{tab:ablation_nfe}). The \textbf{\#Params} exclude the neural codec or vocoder component, which is non-trainable. The best and second best are \textbf{bold} and \underline{underlined}, respectively.}
\resizebox{0.8\textwidth}{!}{%
\begin{tabular}{l|c|c|ccccccc}
\toprule[1.25pt]
\textbf{Model} & \textbf{\#Params} $\downarrow$ & \textbf{NFE} & \textbf{RTF} $\downarrow$ & \textbf{UTMOS} $\uparrow$ & \textbf{WER} $\downarrow$ & \textbf{SIM-O} $\uparrow$ & \textbf{RMSE}$_{\boldsymbol{F0}}$ $\downarrow$ & \textbf{RMSE}$_{\boldsymbol{E}}$ $\downarrow$ \\
\midrule
% Spark-TTS \citep{spark-tts} & 507M & - & 1.062 & \textbf{4.31} & 0.10 & \textbf{0.57} \\
VoiceCraft & 830M & - &  1.70 & 3.55 & 0.18 & 0.51 & 17.22 & 0.010 \\
VALL-E & 594M & - & 0.86 & 3.68 & 0.19 & 0.40 & 21.66 & 0.020 \\
NaturalSpeech 2 & 378M & 200 & 1.66 & 2.38 & 0.09 & 0.31 & 15.62 & 0.020 \\
F5-TTS & 336M & 32 & 0.26 & 3.72 & 0.09 & \underline{0.66} & 12.66 & 0.010 \\
OZSpeech & 145M & 1 & \textbf{0.03} & 3.15 & \textbf{0.05} & 0.40 & 11.96 & 0.010 \\
MaskGCT & 1.43B & 50 + 45$^\dagger$ & 0.46 & 3.83 & 0.09 & \textbf{0.67} & 14.33 & \textbf{0.007} \\
\midrule
\rowcolor{aliceblue}
 & & 4 & \textbf{0.03} & 3.34 & \underline{0.06}	& 0.43 & 8.31 & \textbf{0.007}\\
\rowcolor{aliceblue}
\multirow{-2}{*}{\textbf{DiFlow-TTS-Small (Ours)}} & \multirow{-2}{*}{\textbf{122M}} & 16 & 0.05 & \textbf{3.89} & \textbf{0.05}	& 0.45 & 8.58 & \underline{0.008} \\
% \cdashline{1-7}\noalign{\vskip\belowrulesep}
\midrule
\rowcolor{aliceblue}
& & 4 & \textbf{0.03} & 3.31 & \textbf{0.05}	& 0.44 & \underline{8.05} & \textbf{0.007}\\
\rowcolor{aliceblue}
\multirow{-2}{*}{\textbf{DiFlow-TTS (Ours)}} & \multirow{-2}{*}{164M} & 16 & \underline{0.07} & \underline{3.86} & \textbf{0.05}	& 0.45 & \textbf{7.96} & \textbf{0.007} \\
% \rowcolor{aliceblue}
% & & 16 & 0.80 & \underline{3.86} & \textbf{0.05}	& 0.45\\
\bottomrule[1.25pt]
\multicolumn{9}{l}{$\dagger$ MaskGCT is a two-stage system that first predicts masked semantic tokens, then uses them to infer masked acoustic tokens.} \\
\end{tabular}
}
\label{tab:latency-results}
\end{table*}
% \begin{table*}
% \centering
% \resizebox{\textwidth}{!}{%
% \begin{tabular}{l|cccccccc}
% \toprule[1.25pt]
% \multirow{3}{*}{\textbf{Model}} & \multirow{3}{*}{\textbf{UTMOS} ($\uparrow$)}& \multirow{3}{*}{\textbf{WER} ($\downarrow$)}& \multicolumn{2}{c}{\textbf{SPK-SIM}} & \multicolumn{2}{c}{\textbf{F0}} & \multicolumn{2}{c}{\textbf{Energy}} \\
% \cmidrule(lr){4-5} \cmidrule(lr){6-7} \cmidrule(lr){8-9}
%  &  &  &\textbf{SIM-O} ($\uparrow$) & \textbf{SIM-R} ($\uparrow$) & \textbf{Accuracy} ($\uparrow$) & \textbf{RMSE} ($\downarrow$)& \textbf{Accuracy} ($\uparrow$)& \textbf{RMSE} ($\downarrow$) \\
% \midrule
% \rowcolor{aliceblue}
% \textbf{DiFlow-TTS} & \underline{3.978}	& \textbf{0.048}	& \textbf{0.454}	& \textbf{0.536}	& \textbf{0.884}	& \textbf{7.972}	& \textbf{0.735}	& \textbf{0.007} \\
% - w/o Attribute Embedding & \textbf{3.983}	& 0.060	& \underline{0.444}	& 0.524	& 0.869	& 9.289	& 0.712	& \underline{0.008} \\
% - w/o Speaker Embedding & 3.902	& \underline{0.057}	& 0.378	& 0.461	& 0.681	& 20.868    & 0.615	& 0.010 \\
% - w/o Content Embedding & 3.077	& 0.063	& 0.333	& 0.416	& 0.867	& 8.878	& 0.698	& \underline{0.008} \\
% - w/o Multi-head Prediction & 3.939	& \underline{0.057}	& 0.442	& \underline{0.525}	& \underline{0.876}	& \underline{8.474}	& \underline{0.726}	& \textbf{0.007} \\
% \bottomrule[1.25pt]
% \end{tabular}
% }
% \caption{Ablation study results showing the effect of removing each component from the DiFlow-TTS, with NFE set to 128. The best and second best are \textbf{bold} and \underline{underlined}.}
% \label{tab:ablation_components}
% \end{table*}

\begin{table*}[!httb]
\centering
\caption{Ablation study results showing the effect of removing each component from the DiFlow-TTS, with NFE set to 128. The best and second best are \textbf{bold} and \underline{underlined}, respectively.}
\resizebox{0.8\textwidth}{!}{%
\begin{tabular}{l|ccccccc}
\toprule[1.25pt]
\multirow{3}{*}{\textbf{Model}} & \multirow{3}{*}{\textbf{UTMOS} $\uparrow$}& \multirow{3}{*}{\textbf{WER} $\downarrow$} & \multirow{3}{*}{\textbf{SIM-O} $\uparrow$} & \multicolumn{2}{c}{\textbf{F0}} & \multicolumn{2}{c}{\textbf{Energy}} \\
\cmidrule(lr){5-6} \cmidrule(lr){7-8}
 &  &  & & \textbf{Accuracy} $\uparrow$ & \textbf{RMSE} $\downarrow$& \textbf{Accuracy} $\uparrow$& \textbf{RMSE} $\downarrow$ \\
\midrule
\rowcolor{aliceblue}
\textbf{DiFlow-TTS (Ours)} & \underline{3.978}	& \textbf{0.048}	& \textbf{0.454} & \textbf{0.884}	& \textbf{7.972}	& \textbf{0.735}	& \textbf{0.007} \\
- \emph{w/o} Attribute Embedding & \textbf{3.983}	& 0.060	& \underline{0.444}	& 0.869	& 9.289	& 0.712	& \underline{0.008} \\
- \emph{w/o} Speaker Embedding & 3.902	& \underline{0.057}	& 0.378	& 0.681	& 20.868    & 0.615	& 0.010 \\
- \emph{w/o} Content Embedding & 3.077	& 0.063	& 0.333 & 0.867	& 8.878	& 0.698	& \underline{0.008} \\
- \emph{w/o} Multi-head Prediction & 3.939	& \underline{0.057}	& 0.442	& \underline{0.876}	& \underline{8.474}	& \underline{0.726}	& \textbf{0.007} \\
\bottomrule[1.25pt]
\end{tabular}
}
\label{tab:ablation_components}
\end{table*}
\begin{table*}[t]
    \centering
    % --- Left Side: Table 5 ---
    \begin{minipage}{0.7\textwidth}
        \centering
        \caption{Ablation study on attribute-type embeddings.}
        \resizebox{\textwidth}{!}{%
            \begin{tabular}{l|ccccccc}
            \toprule[1.25pt]
            \multirow{3}{*}{\textbf{Model}} & \multirow{3}{*}{\textbf{UTMOS} $\uparrow$}& \multirow{3}{*}{\textbf{WER} $\downarrow$} & \multirow{3}{*}{\textbf{SIM-O} $\uparrow$} & \multicolumn{2}{c}{\textbf{F0}} & \multicolumn{2}{c}{\textbf{Energy}} \\
            \cmidrule(lr){5-6} \cmidrule(lr){7-8}
             &  &  & & \textbf{Accuracy} $\uparrow$ & \textbf{RMSE} $\downarrow$& \textbf{Accuracy} $\uparrow$& \textbf{RMSE} $\downarrow$ \\
            \midrule
            \rowcolor{aliceblue}
            \textbf{DiFlow-TTS} & \textbf{3.978} & \textbf{0.048} & \textbf{0.454} & \textbf{0.884} & \textbf{7.972} & 0.735 & \textbf{0.007} \\
            - \emph{w/o} Prosody Attribute & 3.918 & 0.062 & 0.445 & 0.846 & 9.547 & 0.734 & 0.007 \\
            - \emph{w/o} Acoustic Attribute & 3.929 & 0.060 & 0.447 & 0.868 & 9.216 & 0.741 & 0.007 \\
            \bottomrule[1.25pt]
            \end{tabular}
        }
        \label{tab:ablation_seg}
    \end{minipage}
    \hfill % Adds horizontal space between the two minipages
    % --- Right Side: Table 6 ---
    \begin{minipage}{0.2\textwidth}
        \centering
        \caption{Performance of UTMOS vs. NFE count.}
        \resizebox{\textwidth}{!}{
            \begin{tabular}{c|cc}
            \toprule[1.25pt]
            \textbf{NFE} & \textbf{RTF} $\downarrow$ & \textbf{UTMOS} $\uparrow$ \\
            \midrule
            1   & 0.022 & 2.904 \\
            2   & 0.025 & 2.908 \\
            4   & 0.031 & 3.313 \\
            8   & 0.043 & 3.698 \\
            16  & 0.066 & 3.864 \\
            32  & 0.112 & 3.923 \\
            64  & 0.207 & 3.958 \\
            128 & 0.394 & 3.978 \\
            \bottomrule[1.25pt]
            \end{tabular}
        }
        \label{tab:ablation_nfe}
    \end{minipage}
\end{table*}
\subsection{Implementation Details}
The PCM has a hidden dimension of 768 and integrates a variance adapter \cite{ren2021fastspeech} for the Duration Predictor with an encoder hidden size of 256, a filter size of 1024, a kernel size of 3, and a dropout rate of 0.5. The Phoneme Encoder consists of 2 FFT layers with 4 attention heads, a hidden size of 256, an output dimension of 768, convolutional filter sizes of 1024 with kernel sizes $[9,1]$, a dropout rate of 0.2, and a maximum sequence length of 5000. The Content Predictor comprises multiple layers, each responsible for extracting textual representations for a corresponding content codebook level. Each layer shares the same FFT-based architecture as the Phoneme Encoder. The FDFD employs a scheduler defined as $\kappa_t = t^2$ and is built on DiT blocks \cite{dit} with a hidden size of 768, 12 layers and 12 attention heads, further enhanced with rotary position embedding (RoPE) \cite{su2024roformer}; the feed-forward network uses a width multiplier of 4. The embedding dimension of the speaker $D_\text{spk}$ is 256.

We train the model on 4$\times$ NVIDIA A100 GPUs for 315K steps with a batch size of 16, using AdamW \cite{loshchilov2018decoupled} with a learning rate of $1 \times 10^{-4}$, weight decay of 0.01, and 200K warm-up steps. For Eq.~\eqref{eq:total_loss}, the overall objective combines duration, content, and denoising losses, weighted by $\lambda_{dur}=0.5$, $\lambda_{c}=1.0$, and $\lambda_{FDFD}=1.0$, respectively.

\subsection{Baselines} 
To ensure a best-effort comparison under \textbf{publicly available checkpoints and code} with standard evaluation settings, we compare against publicly available baselines spanning different modeling paradigms: \textbf{(i)} \textit{Autoregressive models:} VoiceCraft \cite{voicecraft}, VALL-E \cite{valle}; \textbf{(ii)} \textit{Continuous flow matching/diffusion models:} NaturalSpeech 2 \cite{ns2}, F5-TTS \cite{f5tts}, OZSpeech \cite{ozspeech}; \textbf{(iii)} \textit{Masked generative model:} MaskGCT \cite{maskgct}; \textbf{(iv)} \textit{Discrete flow matching model:} DiFlow-TTS. We provide detailed descriptions of the training data, checkpoints, and evaluation settings below:

\begin{itemize}
    \item \textbf{VoiceCraft}: We use the official code and the pre-trained checkpoint\footnote{\url{https://huggingface.co/pyp1/VoiceCraft/blob/main/830M_TTSEnhanced.pth}}, trained on 9K hours of the GigaSpeech dataset \cite{gigaspeech}.
    
    \item \textbf{F5-TTS}: We use samples obtained through communication with the authors of \cite{ozspeech}, reproduced using 500 hours of the LibriTTS dataset. We additionally perform inference using the official code\footnote{\url{https://github.com/SWivid/F5-TTS}} and a pre-trained checkpoint\footnote{\url{https://huggingface.co/SWivid/F5-TTS}} trained on 100K hours of the Emilia dataset \cite{emilia}.
    
    \item \textbf{NaturalSpeech 2}: We use the Amphion toolkit \cite{amphion} and the pre-trained weight\footnote{\url{https://huggingface.co/amphion/naturalspeech2_libritts/tree/main/checkpoint}}, trained on 585 hours of the LibriTTS dataset.
    
    \item \textbf{VALL-E}: We use samples provided through communication with the authors of \cite{ozspeech}. The model is reproduced using the Amphion toolkit\footnote{\url{https://github.com/open-mmlab/Amphion}} \cite{amphion} and trained on 500 hours of the LibriTTS dataset.
    
    \item \textbf{OZSpeech}: We use samples provided through communication with the authors, trained on 500 hours of the LibriTTS dataset.
    
    % \item \textbf{Spark-TTS} \citep{spark-tts}. We use the official code and the pre-trained checkpoint\footnote{\url{https://github.com/SparkAudio/Spark-TTS}}, trained on 100K hours of the VoxBox dataset \citep{spark-tts}.

    \item \textbf{MaskGCT}: We use the official code and the pretrained checkpoint\footnote{\url{https://huggingface.co/amphion/MaskGCT}}, trained on English and Chinese data from Emilia \cite{emilia}, each with 50K hours of speech (totaling 100K hours). Since the baseline requires the total speech length, we use the ground-truth duration during inference.
\end{itemize}

\subsection{Datasets}
\subsubsection{Training}
We use a 470-hour subset of the LibriTTS dataset \cite{libritts}, which comprises multi-speaker English audio recordings, to train our method. We preprocess this dataset for training as follows:

The silent segments at the beginning and end of each utterance are removed. We retain audio clips ranging from 1.0 to 16.6 seconds in duration that contain utterances with more than three words. Speech is sampled at 16 kHz and tokenized using FACodec \cite{ns3} at 80 tokens/s for ground-truth representations, which include a speaker embedding and six sequences of discrete tokens: one for prosody, two for content, and three for acoustic details. To obtain the ground truth of the text at the phoneme-level and the corresponding discrete speech tokens, we use the Montreal Forced Aligner (MFA) \cite{mfa} to align each speech with its target transcription, producing the duration of each phoneme in the speech. We then multiply these durations by 80, which represents the number of tokens per second in FACodec, to determine the number of speech tokens corresponding to each phoneme.
\subsubsection{Evaluation}
We follow the evaluation setup used in OZSpeech \cite{ozspeech} to assess our model on the LibriSpeech test-clean set \cite{librispeech}. This test set consists of 2.2 hours, each ranging from 4 to 10 seconds in length. We perform the cross-sentence generation task using two speech samples from the same speaker: one sample serves as the speech prompt, and the other provides the target text content to be synthesized.

\subsection{Evaluation Metrics}
To evaluate model performance, we employ a suite of \textbf{objective evaluation metrics} that assess multiple aspects of speech synthesis, including naturalness and speech quality (UTMOS), speaker similarity (SIM-O), content accuracy (WER), prosody accuracy and error (Pitch/Energy-Accuracy, Pitch/Energy-RMSE), and inference latency (RTF). Specifically, \ding{182} UTMOS \cite{utmos} is a deep learning-based framework that predicts Mean Opinion Scores (MOS) to estimate speech naturalness and overall quality; \ding{183} for speaker similarity (SIM-O), we compute the cosine similarity between speaker embeddings extracted using a WavLM-TDNN speaker verification model \cite{chen2022wavlm}, applied to the synthesized speech and the reference prompt; \ding{184} for the word error rate (WER), we used the HuBERT-large model \cite{hsu2021hubert} (pre-trained on Libri-Light \cite{librilight}, and fine-tuned on 960h of LibriSpeech \cite{librispeech}) to transcribe the synthesized speech; \ding{185} for prosody accuracy and error, we evaluate the alignment between synthesized speech and the reference prompt in terms of pitch (F0) and energy contours, where, following PromptTTS \cite{prompttts} and TextrolSpeech \cite{textrolspeech}, F0 and energy values are discretized into three relative tiers (high, normal, and low) based on their mean values to compute tier-level accuracy (Pitch/Energy-Accuracy), and Root Mean Square Error (RMSE) is further calculated to quantify deviations in F0 (Pitch-RMSE) and energy (Energy-RMSE) between synthesized outputs and reference prompts; \ding{186} for Real-Time Factor (RTF), we measure inference efficiency by quantifying the time required to generate one second of speech, with all RTF evaluations conducted in a fully end-to-end configuration on a single NVIDIA A100 80GB GPU.

Complementing these objective measures, we perform a \textbf{subjective evaluation} following the Mean Opinion Score (MOS) protocol. Specifically, 30 listeners rate the synthesized speech on a scale from 1 to 5 with respect to  naturalness, intelligibility, and speaker similarity to the speech prompt.

\section{Experimental Results}\label{sec:exp_results}
\subsection{Main Results}
\subsubsection{Comparison Results}
Table~\ref{tab:overall-results-3s} presents the performance of DiFlow-TTS with 128 function evaluations (NFE) using 3-second audio prompts, compared to baseline methods. For naturalness and speech quality as measured by UTMOS, DiFlow-TTS achieves the strongest performance despite being trained on only 470 hours of speech data, which is significantly smaller (by a factor of 1.1$\times$ to 212.8$\times$) than other baselines, highlighting the strength of our FDFD module in capturing prosodic and acoustic nuances even under limited data conditions. For content accuracy, DiFlow-TTS, along with OZSpeech, achieves the best performance in terms of WER, demonstrating the effectiveness of our method in producing speech with accurate linguistic content. For speaker similarity, DiFlow-TTS offers no clear advantage over baselines, likely due to its simple speaker conditioning in DiT blocks, which could be improved with more advanced strategies. \textbf{We highlight this limitation as a promising direction for future work beyond the scope of this study}. For prosody reconstruction, DiFlow-TTS outperforms across all metrics, with the sole exception of energy accuracy, where it trails MaskGCT by only 0.02, despite MaskGCT being trained on significantly more data (100K hours). These findings further confirm the ability of the FDFD module to model fine-grained prosodic attributes with high fidelity. To gain further insight into speech quality, we report the subjective MOS evaluations in Table~\ref{tab:mos-results}. Overall, DiFlow-TTS consistently outperforms other methods across every MOS dimension, providing strong evidence of its well-balanced performance in generating natural and intelligible speech with high speaker similarity. It is worth noting that even though DiFlow-TTS ranks third on SIM-O (see Table \ref{tab:overall-results-3s}), an embedding-space proxy that may penalize artifacts inaudible to humans, it best captures the perceptual identity cues (e.g., pitch, timbre, prosody) that human listeners value, indicating superior speaker faithfulness where it matters most. These results are especially notable given the model’s training data efficiency. For the other MOS metrics, the results are consistent with the findings in Table \ref{tab:overall-results-3s}.

\subsubsection{Model Size \& Latency Analysis}
Table \ref{tab:latency-results} compares the model size and latency between DiFlow-TTS and the baselines for the 3-second audio prompt setting. The RTF metric, measured in seconds, shows that all baselines except OZSpeech and our method experience latency in the order of hundreds of milliseconds. To highlight efficiency, we further construct a  smaller variant of DiFlow-TTS (denoted as \textit{DiFlow-TTS-Small}) by reducing the number of attention heads $(12 \rightarrow 8)$ and DiT layers $(12 \rightarrow 8)$, resulting in a 122M-parameter model. This small variant achieves the best results in both speed and size. For comparison, OZSpeech, optimized for the 1-NFE setting, is expected to perform strongly in a strictly one-step regime. However, despite achieving the same RTF (0.03) as DiFlow-TTS-Small with 4 NFEs, it exhibits significantly lower performance across all metrics except WER; meanwhile, DiFlow-TTS with 4 NFEs (RTF = 0.03) also attains performance comparable to our small variant. Furthermore, DiFlow-TTS-Small with 16 NFEs (RTF = 0.05) achieves competitive performance in naturalness, intelligibility, speaker similarity, and prosody error while being only marginally slower than OZSpeech (by 0.02s in RTF) yet 5.2$\times$ to 34.0$\times$ faster than the other baselines, with a model size 1.2$\times$ to 11.7$\times$ smaller than all baselines. The results of DiFlow-TTS with 16 NFEs further reinforce these findings, demonstrating a strong balance between model size, speed, and high-quality speech.
\begin{figure*}[!httb]
\centering
\begin{minipage}{0.32\textwidth}
    \centering
    \includegraphics[width=\linewidth]{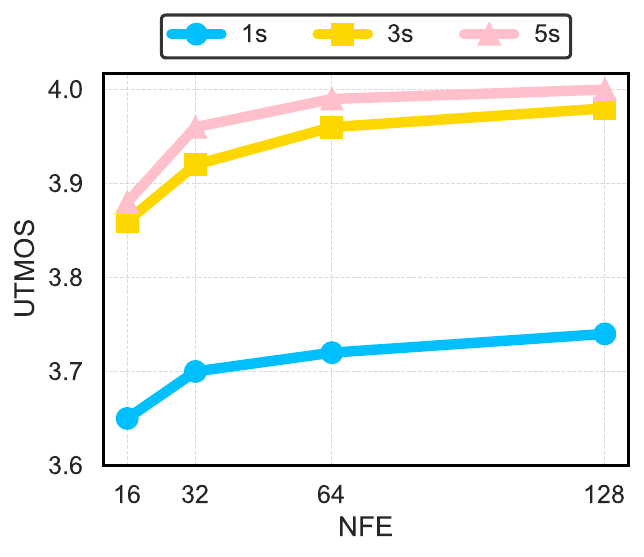}
    \captionof{figure}{UTMOS vs. NFE for different prompt durations.}
    \label{fig:utmos_nfe}
\end{minipage}
\hfill
\begin{minipage}{0.63\textwidth}
    \centering
    \includegraphics[width=\linewidth]{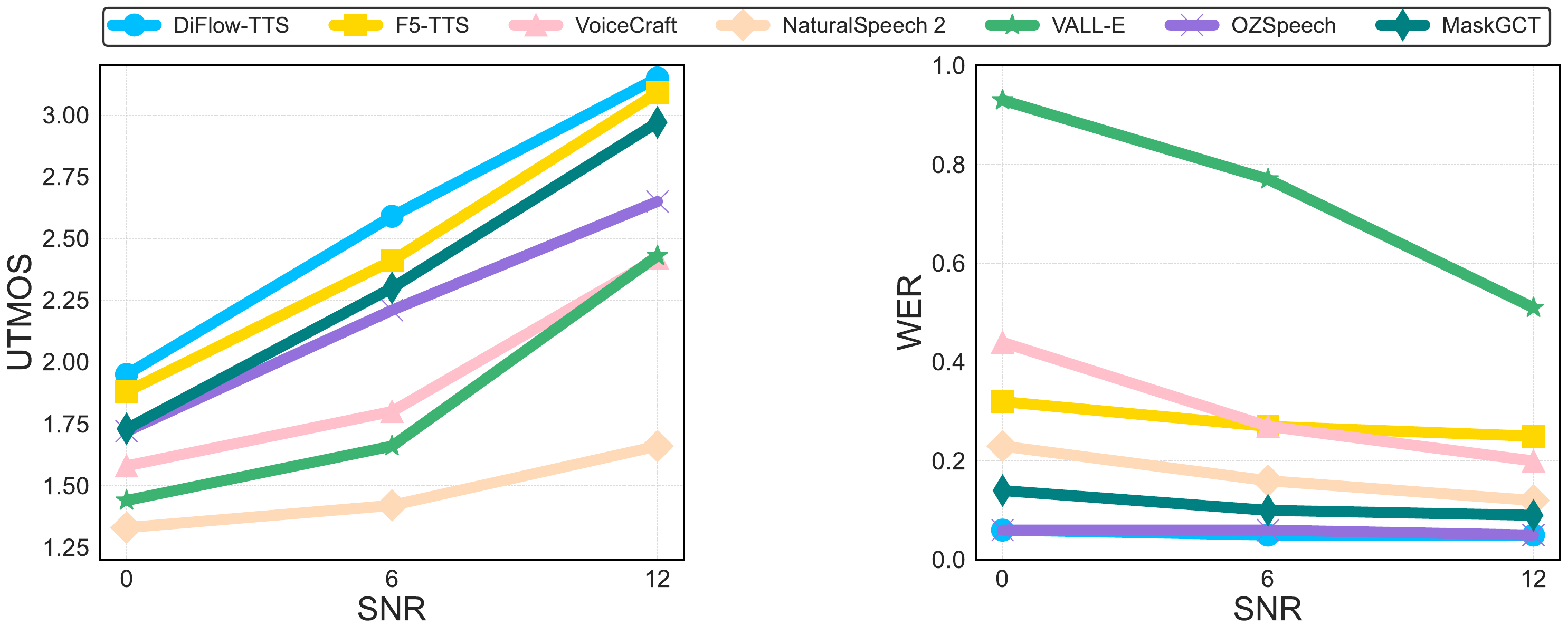}
    \captionof{figure}{Model sensitivity under varying SNR levels in terms of UTMOS (left) and WER (right).}
    \label{fig:snr_robustness}
\end{minipage}
\vspace{-0.3cm}
\end{figure*}
% DiFlow-TTS achieves a RTF of 0.066 and a model size of 164M parameters with 16 NFE, ranking second in both speed and size. In comparison, OZSpeech, with only 1 NFE, is marginally faster (by 0.04s) and slightly smaller (144M parameters), but it exhibits significantly lower UTMOS and speaker similarity scores. Compared to other baselines, DiFlow-TTS is 3.9$\times$ to 25.8$\times$ faster in terms of RTF and 2$\times$ to 5.1$\times$ smaller in model size, while still maintaining competitive performance in speech naturalness, intelligibility, and pronunciation accuracy. These results demonstrate that DiFlow-TTS strikes a strong balance between speed, compactness, and the quality of the speech.

\subsection{Ablation Studies and Analyses}
\subsubsection{Effect of Each Component}
To assess the impact of each component in DiFlow-TTS, we perform an ablation study by systematically removing or modifying key elements: \ding{182} removing the attribute-type embeddings used to distinguish prosody, content, and acoustic streams; \ding{183} excluding the speaker embedding from the conditioning process (i.e., not injecting it into the DiT blocks); \ding{184} disabling the use of content embeddings in the FDFD module; and \ding{185} replacing the multi-head prediction architecture with a single-head prediction. As shown in Table~\ref{tab:ablation_components}, we observe a slight degradation in all metrics except UTMOS when the attribute-type embeddings are removed. This suggests that while these embeddings enhance overall fidelity and prosody modeling, they may introduce minor redundancies that subtly affect perceived naturalness. A more pronounced decline in speaker similarity and prosody-related metrics is observed when speaker embedding is excluded from the FDFD module. This highlights that prosody is not only content-dependent but also strongly influenced by speaker identity; without speaker conditioning, the FDFD module produces extraneous prosodic variations, resulting in reduced speaker adaptation and overall synthesis quality. When content embeddings from the PCM branch are removed from FDFD, we observe substantial degradation across metrics related to naturalness and speaker similarity. This demonstrates the critical role of content embeddings in conditioning FDFD to generate appropriate prosody and support speaker adaptation. Lastly, replacing the multi-head prediction mechanism with a single-head alternative leads to minor performance drops across all metrics, indicating that the multi-head design enhances prediction diversity and robustness in prosody and acoustic modeling.
\subsubsection{Attribute-type Embeddings}
We conduct a finer-grained ablation to isolate the contribution of each attribute-specific embedding. The additional results are reported in Table \ref{tab:ablation_seg}. We observe that removing either the prosody attribute embedding or the acoustic attribute embedding leads to noticeable performance degradation, particularly in naturalness and metrics closely tied to these attributes. Attribute-type embeddings guide the model in distinguishing which attribute each token corresponds to when feeding them into the DiT. When an attribute embedding is removed, the model is forced to treat all attributes uniformly, which results in entanglement and reduced synthesis quality.
\subsubsection{Effect of NFE}
We investigate the impact of varying NFE from 1 to 128 on DiFlow-TTS performance to explore the trade-off between inference efficiency and synthesis quality, as presented in Table \ref{tab:ablation_nfe}. Increasing NFE markedly improves UTMOS, indicating that the FDFD module benefits from additional refinement steps to generate more natural speech. In particular, performance stabilizes around 32 NFE, with optimal audio quality observed at 64 NFE, and only marginal improvements beyond this point. Although RTF naturally increases with NFE, the overall latency remains competitive (see Table \ref{tab:latency-results}). As reported in the Table~\ref{tab:overall-results-3s}, we show the \textbf{best performance at NFE = 128}; lower NFE values can be selected based on the user's latency and quality requirements.

\vspace{-0.2cm}
\subsubsection{Prompt Duration Analysis}
To gain further insight into model behavior, we report the relationship between UTMOS, which strongly correlates with human perceptual evaluations, and NFE across different prompt durations, as illustrated in Figure~\ref{fig:utmos_nfe}. In general, longer prompts lead to higher UTMOS scores, indicating improved reconstruction quality and a greater sensitivity of the model to prompt length. Additionally, increasing the NFE from 16 to 128 consistently improves performance for all prompt durations. In particular, the highest performance is achieved with a 5-second prompt and an NFE of 128.
% We refer readers to Appendix \ref{app:additional_analysis} for detailed comparisons of DiFlow-TTS against baselines
% under varying prompt lengths.
\vspace{-0.2cm}
\subsubsection{Noisy Prompt Analysis}
\vspace{-0.2cm}
% \input{figures/snr_utmos_wer}
% \begin{wrapfigure}{r}{0.5\textwidth}
%     \centering
%     \includegraphics[width=\linewidth]{figures/snr_robustness.pdf}
%     \caption{Caption.}
% \end{wrapfigure}
We analyze the robustness of DiFlow-TTS under noisy audio prompts using UTMOS and WER, a challenging scenario since most models are trained on clean speech. The noisy prompts are generated from the LibriSpeech test-clean set with additive noise augmentation. As shown in Figure \ref{fig:snr_robustness}, all models are highly sensitive to noise, showing sharp degradation in both UTMOS (left) and WER (right) as the prompt SNR decreases (see Table \ref{tab:overall-results-3s} for clean-prompt reference). DiFlow-TTS, however, consistently achieves the highest UTMOS across all noise levels, demonstrating its ability to synthesize high-fidelity speech under noisy conditions. For WER, it shows little to no degradation across SNR levels, a trend also observed in OZSpeech, while other baselines suffer significant performance drops.
\section{Conclusion}
%This paper introduces DiFlow-TTS, a novel zero-shot TTS system that leverages discrete flow matching, enabling high-quality speech synthesis in discrete settings. By combining a PCM for accurate content modeling with a FDFD that separately models prosody and acoustic attributes through attribute-specific heads, DiFlow-TTS delivers strong performance in naturalness, intelligibility, prosody, and inference speed, as confirmed by comprehensive objective and subjective evaluations. These results establish DiFlow-TTS as a compelling solution for efficient, high-fidelity zero-shot speech synthesis, well-suited to resource-constrained and latency-sensitive applications, and highlight discrete flow models as a promising direction for future generative speech research.
In this paper, we introduce \textbf{DiFlow-TTS}, a novel zero-shot TTS framework that brings discrete flow matching into the domain of speech generation. By combining the PCM for accurate content modeling with the FDFD that separately models prosody and acoustic attributes, DiFlow-TTS achieves strong performance in naturalness, intelligibility, prosody, and inference efficiency, while exhibiting limitations in speaker similarity as reflected by comprehensive objective and subjective evaluations. These results establish DiFlow-TTS as a compelling solution for efficient, high-fidelity zero-shot speech synthesis, well-suited to resource-constrained and latency-sensitive applications, and highlight discrete flow models as a promising direction for future generative speech research.

\vspace{-0.2cm}
\section{Limitations and Future Works}
We demonstrated the viability of discrete flow matching for high-quality speech synthesis; however, our proposed method still exhibits several limitations. In particular, the current design \textbf{does not fully meet expectations in terms of voice cloning quality}; we acknowledge that our current strategy for speaker conditioning is not yet optimal. Since our framework relies on FACodec~\cite{ns3}, which explicitly disentangles speech attributes and speaker identity, integrating it effectively requires additional mechanisms to inject speaker information. In contrast, codecs such as EnCodec~\cite{encodec} embed speaker information directly within their vector quantizers, enabling models to implicitly learn speaker characteristics.

In future work, we plan to extend our framework to \textbf{support alternative codec models} or to \textbf{replace global AdaLN conditioning with the in-context conditioning mechanism specifically designed for timbre embeddings}, allowing the model to capture local speaker characteristics more effectively. Developing more robust methods for reproducing speaker timbre from real-world audio in discrete settings, particularly in zero-shot TTS scenarios, remains an open research direction.
\section{Generative AI Use Disclosure}
The authors used Generative AI tools exclusively for manuscript refinement, including polishing and improving clarity, grammar, and sentence structure. All AI-assisted outputs were carefully reviewed and edited by the authors, who take full responsibility for the content of the current version.

\bibliographystyle{IEEEtran}
\bibliography{main}

@inproceedings{f5tts,
    title = "F5-{TTS}: A Fairytaler that Fakes Fluent and Faithful Speech with Flow Matching",
    author = "Chen, Yushen  and
      Niu, Zhikang  and
      Ma, Ziyang  and
      Deng, Keqi  and
      Wang, Chunhui  and
      JianZhao, JianZhao  and
      Yu, Kai  and
      Chen, Xie",
    editor = "Che, Wanxiang  and
      Nabende, Joyce  and
      Shutova, Ekaterina  and
      Pilehvar, Mohammad Taher",
    booktitle = "Proceedings of the 63rd Annual Meeting of the Association for Computational Linguistics (Volume 1: Long Papers)",
    month = jul,
    year = "2025",
    address = "Vienna, Austria",
    publisher = "Association for Computational Linguistics",
    url = "https://aclanthology.org/2025.acl-long.313/",
    doi = "10.18653/v1/2025.acl-long.313",
    pages = "6255--6271",
    ISBN = "979-8-89176-251-0"
}

@inproceedings{voicecraft,
    title = "{V}oice{C}raft: Zero-Shot Speech Editing and Text-to-Speech in the Wild",
    author = "Peng, Puyuan  and
      Huang, Po-Yao  and
      Li, Shang-Wen  and
      Mohamed, Abdelrahman  and
      Harwath, David",
    editor = "Ku, Lun-Wei  and
      Martins, Andre  and
      Srikumar, Vivek",
    booktitle = "Proceedings of the 62nd Annual Meeting of the Association for Computational Linguistics (Volume 1: Long Papers)",
    month = aug,
    year = "2024",
    address = "Bangkok, Thailand",
    publisher = "Association for Computational Linguistics",
    url = "https://aclanthology.org/2024.acl-long.673",
    doi = "10.18653/v1/2024.acl-long.673",
    pages = "12442--12462",
}

@inproceedings{
ns2,
title={NaturalSpeech 2: Latent Diffusion Models are Natural and Zero-Shot Speech and Singing Synthesizers},
author={Kai Shen and Zeqian Ju and Xu Tan and Eric Liu and Yichong Leng and Lei He and Tao Qin and sheng zhao and Jiang Bian},
booktitle={The Twelfth International Conference on Learning Representations},
year={2024},
url={https://openreview.net/forum?id=Rc7dAwVL3v}
}

@ARTICLE{valle,
  author={Chen, Sanyuan and Wang, Chengyi and Wu, Yu and Zhang, Ziqiang and Zhou, Long and Liu, Shujie and Chen, Zhuo and Liu, Yanqing and Wang, Huaming and Li, Jinyu and He, Lei and Zhao, Sheng and Wei, Furu},
  journal={IEEE Transactions on Audio, Speech and Language Processing}, 
  title={Neural Codec Language Models are Zero-Shot Text to Speech Synthesizers}, 
  year={2025},
  volume={},
  number={},
  pages={1-15},
  doi={10.1109/TASLPRO.2025.3530270}}

@inproceedings{dfm,
 author = {Gat, Itai and Remez, Tal and Shaul, Neta and Kreuk, Felix and Chen, Ricky T. Q. and Synnaeve, Gabriel and Adi, Yossi and Lipman, Yaron},
 booktitle = {Advances in Neural Information Processing Systems},
 editor = {A. Globerson and L. Mackey and D. Belgrave and A. Fan and U. Paquet and J. Tomczak and C. Zhang},
 pages = {133345--133385},
 publisher = {Curran Associates, Inc.},
 title = {Discrete Flow Matching},
 url = {https://proceedings.neurips.cc/paper_files/paper/2024/file/f0d629a734b56a642701bba7bc8bb3ed-Paper-Conference.pdf},
 volume = {37},
 year = {2024}
}

@inproceedings{ozspeech,
    title = "{OZS}peech: One-step Zero-shot Speech Synthesis with Learned-Prior-Conditioned Flow Matching",
    author = "Hieu, Nghia Huynh Nguyen  and
      Nguyen, Ngoc Son  and
      Dang, Huynh Nguyen  and
      Vo, Thieu  and
      Hy, Truong-Son  and
      Nguyen, Van",
    editor = "Che, Wanxiang  and
      Nabende, Joyce  and
      Shutova, Ekaterina  and
      Pilehvar, Mohammad Taher",
    booktitle = "Proceedings of the 63rd Annual Meeting of the Association for Computational Linguistics (Volume 1: Long Papers)",
    month = jul,
    year = "2025",
    address = "Vienna, Austria",
    publisher = "Association for Computational Linguistics",
    url = "https://aclanthology.org/2025.acl-long.1043/",
    pages = "21500--21517",
    ISBN = "979-8-89176-251-0"
}

@article{encodec,
title={High Fidelity Neural Audio Compression},
author={Alexandre D{\'e}fossez and Jade Copet and Gabriel Synnaeve and Yossi Adi},
journal={Transactions on Machine Learning Research},
issn={2835-8856},
year={2023},
url={https://openreview.net/forum?id=ivCd8z8zR2},
note={Featured Certification, Reproducibility Certification}
}

@InProceedings{ns3,
  title = 	 {{N}atural{S}peech 3: Zero-Shot Speech Synthesis with Factorized Codec and Diffusion Models},
  author =       {Ju, Zeqian and Wang, Yuancheng and Shen, Kai and Tan, Xu and Xin, Detai and Yang, Dongchao and Liu, Eric and Leng, Yichong and Song, Kaitao and Tang, Siliang and Wu, Zhizheng and Qin, Tao and Li, Xiangyang and Ye, Wei and Zhang, Shikun and Bian, Jiang and He, Lei and Li, Jinyu and Zhao, Sheng},
  booktitle = 	 {Proceedings of the 41st International Conference on Machine Learning},
  pages = 	 {22605--22623},
  year = 	 {2024},
  editor = 	 {Salakhutdinov, Ruslan and Kolter, Zico and Heller, Katherine and Weller, Adrian and Oliver, Nuria and Scarlett, Jonathan and Berkenkamp, Felix},
  volume = 	 {235},
  series = 	 {Proceedings of Machine Learning Research},
  month = 	 {21--27 Jul},
  publisher =    {PMLR},
  url = 	 {https://proceedings.mlr.press/v235/ju24b.html}
}

@article{spark-tts,
  title={Spark-tts: An efficient llm-based text-to-speech model with single-stream decoupled speech tokens},
  author={Wang, Xinsheng and Jiang, Mingqi and Ma, Ziyang and Zhang, Ziyu and Liu, Songxiang and Li, Linqin and Liang, Zheng and Zheng, Qixi and Wang, Rui and Feng, Xiaoqin and others},
  journal={arXiv preprint arXiv:2503.01710},
  year={2025}
}

@inproceedings{melle,
    title = "Autoregressive Speech Synthesis without Vector Quantization",
    author = "Meng, Lingwei  and
      Zhou, Long  and
      Liu, Shujie  and
      Chen, Sanyuan  and
      Han, Bing  and
      Hu, Shujie  and
      Liu, Yanqing  and
      Li, Jinyu  and
      Zhao, Sheng  and
      Wu, Xixin  and
      Meng, Helen M.  and
      Wei, Furu",
    editor = "Che, Wanxiang  and
      Nabende, Joyce  and
      Shutova, Ekaterina  and
      Pilehvar, Mohammad Taher",
    booktitle = "Proceedings of the 63rd Annual Meeting of the Association for Computational Linguistics (Volume 1: Long Papers)",
    month = jul,
    year = "2025",
    address = "Vienna, Austria",
    publisher = "Association for Computational Linguistics",
    url = "https://aclanthology.org/2025.acl-long.65/",
    pages = "1287--1300",
    ISBN = "979-8-89176-251-0"
}

@inproceedings{
lee2025dittotts,
title={Di{TT}o-{TTS}: Diffusion Transformers for Scalable Text-to-Speech without Domain-Specific Factors},
author={Keon Lee and Dong Won Kim and Jaehyeon Kim and Seungjun Chung and Jaewoong Cho},
booktitle={The Thirteenth International Conference on Learning Representations},
year={2025},
url={https://openreview.net/forum?id=hQvX9MBowC}
}

@INPROCEEDINGS{matcha-tts,
  author={Mehta, Shivam and Tu, Ruibo and Beskow, Jonas and Sz\'ekely, \'Eva and Henter, Gustav Eje},
  booktitle={ICASSP 2024 - 2024 IEEE International Conference on Acoustics, Speech and Signal Processing (ICASSP)}, 
  title={Matcha-TTS: A Fast TTS Architecture with Conditional Flow Matching}, 
  year={2024},
  volume={},
  number={},
  pages={11341-11345},
  doi={10.1109/ICASSP48485.2024.10448291}
}

@inproceedings{
vq-wav2vec,
title={vq-wav2vec: Self-Supervised Learning of Discrete Speech Representations},
author={Alexei Baevski and Steffen Schneider and Michael Auli},
booktitle={International Conference on Learning Representations},
year={2020},
url={https://openreview.net/forum?id=rylwJxrYDS}
}

@ARTICLE{diffsound,
  author={Yang, Dongchao and Yu, Jianwei and Wang, Helin and Wang, Wen and Weng, Chao and Zou, Yuexian and Yu, Dong},
  journal={IEEE/ACM Transactions on Audio, Speech, and Language Processing}, 
  title={Diffsound: Discrete Diffusion Model for Text-to-Sound Generation}, 
  year={2023},
  volume={31},
  number={},
  pages={1720-1733},
  doi={10.1109/TASLP.2023.3268730}}

@ARTICLE{instructtts,
  author={Yang, Dongchao and Liu, Songxiang and Huang, Rongjie and Weng, Chao and Meng, Helen},
  journal={IEEE/ACM Transactions on Audio, Speech, and Language Processing}, 
  title={InstructTTS: Modelling Expressive TTS in Discrete Latent Space With Natural Language Style Prompt}, 
  year={2024},
  volume={32},
  number={},
  pages={2913-2925},
  doi={10.1109/TASLP.2024.3402088}}

@INPROCEEDINGS{dctts,
  author={Wu, Zhichao and Li, Qiulin and Liu, Sixing and Yang, Qun},
  booktitle={ICASSP 2024 - 2024 IEEE International Conference on Acoustics, Speech and Signal Processing (ICASSP)}, 
  title={DCTTS: Discrete Diffusion Model with Contrastive Learning for Text-to-Speech Generation}, 
  year={2024},
  volume={},
  number={},
  pages={11336-11340},
  doi={10.1109/ICASSP48485.2024.10447661}}

@INPROCEEDINGS{emilia,
  author={He, Haorui and Shang, Zengqiang and Wang, Chaoren and Li, Xuyuan and Gu, Yicheng and Hua, Hua and Liu, Liwei and Yang, Chen and Li, Jiaqi and Shi, Peiyang and Wang, Yuancheng and Chen, Kai and Zhang, Pengyuan and Wu, Zhizheng},
  booktitle={2024 IEEE Spoken Language Technology Workshop (SLT)}, 
  title={Emilia: An Extensive, Multilingual, and Diverse Speech Dataset For Large-Scale Speech Generation}, 
  year={2024},
  volume={},
  number={},
  pages={885-890},
  doi={10.1109/SLT61566.2024.10832365}}

@article{hubert,
author = {Hsu, Wei-Ning and Bolte, Benjamin and Tsai, Yao-Hung Hubert and Lakhotia, Kushal and Salakhutdinov, Ruslan and Mohamed, Abdelrahman},
title = {HuBERT: Self-Supervised Speech Representation Learning by Masked Prediction of Hidden Units},
year = {2021},
issue_date = {2021},
publisher = {IEEE Press},
volume = {29},
issn = {2329-9290},
url = {https://doi.org/10.1109/TASLP.2021.3122291},
doi = {10.1109/TASLP.2021.3122291},
journal = {IEEE/ACM Trans. Audio, Speech and Lang. Proc.},
month = oct,
pages = {3451–3460},
numpages = {10}
}

@inproceedings{zuo2025enhancing,
  title={Enhancing expressive voice conversion with discrete pitch-conditioned flow matching model},
  author={Zuo, Jialong and Ji, Shengpeng and Fang, Minghui and Jiang, Ziyue and Cheng, Xize and Yang, Qian and Liu, Wenrui and Zhang, Guangyan and Tu, Zehai and Guo, Yiwen and others},
  booktitle={ICASSP 2025-2025 IEEE International Conference on Acoustics, Speech and Signal Processing (ICASSP)},
  pages={1--5},
  year={2025},
  organization={IEEE}
}

@article{fuest2025maskflow,
  title={Maskflow: Discrete flows for flexible and efficient long video generation},
  author={Fuest, Michael and Hu, Vincent Tao and Ommer, Bj{\"o}rn},
  journal={arXiv preprint arXiv:2502.11234},
  year={2025}
}

@inproceedings{dit,
  title={Scalable diffusion models with transformers},
  author={Peebles, William and Xie, Saining},
  booktitle={Proceedings of the IEEE/CVF international conference on computer vision},
  pages={4195--4205},
  year={2023}
}

@inproceedings{VQ-VAE,
author = {van den Oord, Aaron and Vinyals, Oriol and Kavukcuoglu, Koray},
title = {Neural discrete representation learning},
year = {2017},
isbn = {9781510860964},
publisher = {Curran Associates Inc.},
address = {Red Hook, NY, USA},
booktitle = {Proceedings of the 31st International Conference on Neural Information Processing Systems},
pages = {6309–6318},
numpages = {10},
location = {Long Beach, California, USA},
series = {NIPS'17}
}

@inproceedings{
jia2025ditar,
title={Di{TAR}: Diffusion Transformer Autoregressive Modeling for Speech Generation},
author={Dongya Jia and Zhuo Chen and Jiawei Chen and Chenpeng Du and Jian Wu and Jian Cong and Xiaobin Zhuang and Chumin Li and Zhen Wei and Yuping Wang and Yuxuan Wang},
booktitle={Forty-second International Conference on Machine Learning},
year={2025},
url={https://openreview.net/forum?id=8tRtweTTwv}
}

@article{unicasts, title={UniCATS: A Unified Context-Aware Text-to-Speech Framework with Contextual VQ-Diffusion and Vocoding}, volume={38}, url={https://ojs.aaai.org/index.php/AAAI/article/view/29747}, DOI={10.1609/aaai.v38i16.29747}, abstractNote={The utilization of discrete speech tokens, divided into semantic tokens and acoustic tokens, has been proven superior to traditional acoustic feature mel-spectrograms in terms of naturalness and robustness for text-to-speech (TTS) synthesis. Recent popular models, such as VALL-E and SPEAR-TTS, allow zero-shot speaker adaptation through auto-regressive (AR) continuation of acoustic tokens extracted from a short speech prompt. However, these AR models are restricted to generate speech only in a left-to-right direction, making them unsuitable for speech editing where both preceding and following contexts are provided. Furthermore, these models rely on acoustic tokens, which have audio quality limitations imposed by the performance of audio codec models. In this study, we propose a unified context-aware TTS framework called UniCATS, which is capable of both speech continuation and editing. UniCATS comprises two components, an acoustic model CTX-txt2vec and a vocoder CTX-vec2wav. CTX-txt2vec employs contextual VQ-diffusion to predict semantic tokens from the input text, enabling it to incorporate the semantic context and maintain seamless concatenation with the surrounding context. Following that, CTX-vec2wav utilizes contextual vocoding to convert these semantic tokens into waveforms, taking into consideration the acoustic context. Our experimental results demonstrate that CTX-vec2wav outperforms HifiGAN and AudioLM in terms of speech resynthesis from semantic tokens. Moreover, we show that UniCATS achieves state-of-the-art performance in both speech continuation and editing. Audio samples are available at https://cpdu.github.io/unicats.}, number={16}, journal={Proceedings of the AAAI Conference on Artificial Intelligence}, author={Du, Chenpeng and Guo, Yiwei and Shen, Feiyu and Liu, Zhijun and Liang, Zheng and Chen, Xie and Wang, Shuai and Zhang, Hui and Yu, Kai}, year={2024}, month={Mar.}, pages={17924-17932} }

@inproceedings{
cfm,
title={Flow Matching for Generative Modeling},
author={Yaron Lipman and Ricky T. Q. Chen and Heli Ben-Hamu and Maximilian Nickel and Matthew Le},
booktitle={The Eleventh International Conference on Learning Representations },
year={2023},
url={https://openreview.net/forum?id=PqvMRDCJT9t}
}

@inproceedings{NEURIPS2020_4c5bcfec,
 author = {Ho, Jonathan and Jain, Ajay and Abbeel, Pieter},
 booktitle = {Advances in Neural Information Processing Systems},
 editor = {H. Larochelle and M. Ranzato and R. Hadsell and M.F. Balcan and H. Lin},
 pages = {6840--6851},
 publisher = {Curran Associates, Inc.},
 title = {Denoising Diffusion Probabilistic Models},
 url = {https://proceedings.neurips.cc/paper_files/paper/2020/file/4c5bcfec8584af0d967f1ab10179ca4b-Paper.pdf},
 volume = {33},
 year = {2020}
}

@inproceedings{
song2021scorebased,
title={Score-Based Generative Modeling through Stochastic Differential Equations},
author={Yang Song and Jascha Sohl-Dickstein and Diederik P Kingma and Abhishek Kumar and Stefano Ermon and Ben Poole},
booktitle={International Conference on Learning Representations},
year={2021},
url={https://openreview.net/forum?id=PxTIG12RRHS}
}

@inproceedings{
rectified-flow,
title={Flow Straight and Fast: Learning to Generate and Transfer Data with Rectified Flow},
author={Xingchao Liu and Chengyue Gong and qiang liu},
booktitle={The Eleventh International Conference on Learning Representations },
year={2023},
url={https://openreview.net/forum?id=XVjTT1nw5z}
}

@inproceedings{guan2024reflow,
  title={Reflow-tts: A rectified flow model for high-fidelity text-to-speech},
  author={Guan, Wenhao and Su, Qi and Zhou, Haodong and Miao, Shiyu and Xie, Xingjia and Li, Lin and Hong, Qingyang},
  booktitle={ICASSP 2024-2024 IEEE International Conference on Acoustics, Speech and Signal Processing (ICASSP)},
  pages={10501--10505},
  year={2024},
  organization={IEEE}
}

@inproceedings{yao2025stablevc,
  title={Stablevc: Style controllable zero-shot voice conversion with conditional flow matching},
  author={Yao, Jixun and Yuguang, Yang and Pan, Yu and Ning, Ziqian and Ye, Jianhao and Zhou, Hongbin and Xie, Lei},
  booktitle={Proceedings of the AAAI Conference on Artificial Intelligence},
  volume={39},
  pages={25669--25677},
  year={2025}
}

@InProceedings{pmlr-v235-lou24a,
  title = 	 {Discrete Diffusion Modeling by Estimating the Ratios of the Data Distribution},
  author =       {Lou, Aaron and Meng, Chenlin and Ermon, Stefano},
  booktitle = 	 {Proceedings of the 41st International Conference on Machine Learning},
  pages = 	 {32819--32848},
  year = 	 {2024},
  editor = 	 {Salakhutdinov, Ruslan and Kolter, Zico and Heller, Katherine and Weller, Adrian and Oliver, Nuria and Scarlett, Jonathan and Berkenkamp, Felix},
  volume = 	 {235},
  series = 	 {Proceedings of Machine Learning Research},
  month = 	 {21--27 Jul},
  publisher =    {PMLR},
  url = 	 {https://proceedings.mlr.press/v235/lou24a.html}
}

@inproceedings{
shi2024simplified,
title={Simplified and Generalized Masked Diffusion for Discrete Data},
author={Jiaxin Shi and Kehang Han and Zhe Wang and Arnaud Doucet and Michalis Titsias},
booktitle={The Thirty-eighth Annual Conference on Neural Information Processing Systems},
year={2024},
url={https://openreview.net/forum?id=xcqSOfHt4g}
}

@inproceedings{CampbellYBRJ24,
  author={Andrew Campbell and Jason Yim and Regina Barzilay and Tom Rainforth and Tommi S. Jaakkola},
  title={Generative Flows on Discrete State-Spaces: Enabling Multimodal Flows with Applications to Protein Co-Design},
  year={2024},
  cdate={1704067200000},
  url={https://openreview.net/forum?id=kQwSbv0BR4},
  booktitle={ICML}
}

@inproceedings{
yi2025allatom,
title={All-atom inverse protein folding through discrete flow matching},
author={Kai Yi and Kiarash Jamali and Sjors HW Scheres},
booktitle={Forty-second International Conference on Machine Learning},
year={2025},
url={https://openreview.net/forum?id=8tQdwSCJmA}
}

@inproceedings{qin2025defog,
          title     = {DeFoG: Discrete Flow Matching for Graph Generation},
          author    = {Qin, Yiming and Madeira, Manuel and Thanou, Dorina and Frossard, Pascal},
          booktitle = {Proceedings of the 42nd International Conference on Machine Learning (ICML)},
          year      = {2025},
}

@InProceedings{chang2022maskgit,
    author    = {Chang, Huiwen and Zhang, Han and Jiang, Lu and Liu, Ce and Freeman, William T.},
    title     = {MaskGIT: Masked Generative Image Transformer},
    booktitle = {Proceedings of the IEEE/CVF Conference on Computer Vision and Pattern Recognition (CVPR)},
    month     = {June},
    year      = {2022},
    pages     = {11315-11325}
}

@inproceedings{libritts,
  title     = {LibriTTS: A Corpus Derived from LibriSpeech for Text-to-Speech},
  author    = {Heiga Zen and Viet Dang and Rob Clark and Yu Zhang and Ron J. Weiss and Ye Jia and Zhifeng Chen and Yonghui Wu},
  year      = {2019},
  booktitle = {Interspeech 2019},
  pages     = {1526--1530},
  doi       = {10.21437/Interspeech.2019-2441},
  issn      = {2958-1796},
}

@INPROCEEDINGS{librispeech,
  author={Panayotov, Vassil and Chen, Guoguo and Povey, Daniel and Khudanpur, Sanjeev},
  booktitle={2015 IEEE International Conference on Acoustics, Speech and Signal Processing (ICASSP)}, 
  title={Librispeech: An ASR corpus based on public domain audio books}, 
  year={2015},
  volume={},
  number={},
  pages={5206-5210},
  doi={10.1109/ICASSP.2015.7178964}}

@inproceedings{utmos,
  title     = {{UTMOS: UTokyo-SaruLab System for VoiceMOS Challenge 2022}},
  author    = {Takaaki Saeki and Detai Xin and Wataru Nakata and Tomoki Koriyama and Shinnosuke Takamichi and Hiroshi Saruwatari},
  year      = {2022},
  booktitle = {{Interspeech 2022}},
  pages     = {4521--4525},
  doi       = {10.21437/Interspeech.2022-439},
  issn      = {2958-1796},
}

@misc{prompttts,
      title={PromptTTS: Controllable Text-to-Speech with Text Descriptions}, 
      author={Zhifang Guo and Yichong Leng and Yihan Wu and Sheng Zhao and Xu Tan},
      year={2022},
      eprint={2211.12171},
      archivePrefix={arXiv},
      primaryClass={eess.AS},
      url={https://arxiv.org/abs/2211.12171}, 
}

@INPROCEEDINGS{textrolspeech,
  author={Ji, Shengpeng and Zuo, Jialong and Fang, Minghui and Jiang, Ziyue and Chen, Feiyang and Duan, Xinyu and Huai, Baoxing and Zhao, Zhou},
  booktitle={ICASSP 2024 - 2024 IEEE International Conference on Acoustics, Speech and Signal Processing (ICASSP)}, 
  title={TextrolSpeech: A Text Style Control Speech Corpus with Codec Language Text-to-Speech Models}, 
  year={2024},
  volume={},
  number={},
  pages={10301-10305},
  doi={10.1109/ICASSP48485.2024.10445879}}

@inproceedings{gigaspeech,
  title     = {GigaSpeech: An Evolving, Multi-Domain ASR Corpus with 10,000 Hours of Transcribed Audio},
  author    = {Guoguo Chen and Shuzhou Chai and Guan-Bo Wang and Jiayu Du and Wei-Qiang Zhang and Chao Weng and Dan Su and Daniel Povey and Jan Trmal and Junbo Zhang and Mingjie Jin and Sanjeev Khudanpur and Shinji Watanabe and Shuaijiang Zhao and Wei Zou and Xiangang Li and Xuchen Yao and Yongqing Wang and Zhao You and Zhiyong Yan},
  year      = {2021},
  booktitle = {Interspeech 2021},
  pages     = {3670--3674},
  doi       = {10.21437/Interspeech.2021-1965},
  issn      = {2958-1796},
}

@inproceedings{amphion,
    author={Xueyao Zhang and Liumeng Xue and Yicheng Gu and Yuancheng Wang and Jiaqi Li and Haorui He and Chaoren Wang and Ting Song and Xi Chen and Zihao Fang and Haopeng Chen and Junan Zhang and Tze Ying Tang and Lexiao Zou and Mingxuan Wang and Jun Han and Kai Chen and Haizhou Li and Zhizheng Wu},
    title={Amphion: An Open-Source Audio, Music and Speech Generation Toolkit},
    booktitle={{IEEE} Spoken Language Technology Workshop, {SLT} 2024},
    year={2024}
}

@article{su2024roformer,
  title={Roformer: Enhanced transformer with rotary position embedding},
  author={Su, Jianlin and Ahmed, Murtadha and Lu, Yu and Pan, Shengfeng and Bo, Wen and Liu, Yunfeng},
  journal={Neurocomputing},
  volume={568},
  pages={127063},
  year={2024},
  publisher={Elsevier}
}

@inproceedings{
loshchilov2018decoupled,
title={Decoupled Weight Decay Regularization},
author={Ilya Loshchilov and Frank Hutter},
booktitle={International Conference on Learning Representations},
year={2019},
url={https://openreview.net/forum?id=Bkg6RiCqY7},
}

@inproceedings{mfa,
  title     = {Montreal Forced Aligner: Trainable Text-Speech Alignment Using Kaldi},
  author    = {Michael McAuliffe and Michaela Socolof and Sarah Mihuc and Michael Wagner and Morgan Sonderegger},
  year      = {2017},
  booktitle = {Interspeech 2017},
  pages     = {498--502},
  doi       = {10.21437/Interspeech.2017-1386},
  issn      = {2958-1796},
}

@inproceedings{
ren2021fastspeech,
title={FastSpeech 2: Fast and High-Quality End-to-End Text to Speech},
author={Yi Ren and Chenxu Hu and Xu Tan and Tao Qin and Sheng Zhao and Zhou Zhao and Tie-Yan Liu},
booktitle={International Conference on Learning Representations},
year={2021},
url={https://openreview.net/forum?id=piLPYqxtWuA}
}

@inproceedings{
maskgct,
title={Mask{GCT}: Zero-Shot Text-to-Speech with Masked Generative Codec Transformer},
author={Yuancheng Wang and Haoyue Zhan and Liwei Liu and Ruihong Zeng and Haotian Guo and Jiachen Zheng and Qiang Zhang and Xueyao Zhang and Shunsi Zhang and Zhizheng Wu},
booktitle={The Thirteenth International Conference on Learning Representations},
year={2025},
url={https://openreview.net/forum?id=ExuBFYtCQU}
}

@article{hsu2021hubert,
  title={Hubert: Self-supervised speech representation learning by masked prediction of hidden units},
  author={Hsu, Wei-Ning and Bolte, Benjamin and Tsai, Yao-Hung Hubert and Lakhotia, Kushal and Salakhutdinov, Ruslan and Mohamed, Abdelrahman},
  journal={IEEE/ACM transactions on audio, speech, and language processing},
  volume={29},
  pages={3451--3460},
  year={2021},
  publisher={IEEE}
}

@INPROCEEDINGS{librilight,
  author={J. {Kahn} and M. {Rivière} and W. {Zheng} and E. {Kharitonov} and Q. {Xu} and P. E. {Mazaré} and J. {Karadayi} and V. {Liptchinsky} and R. {Collobert} and C. {Fuegen} and T. {Likhomanenko} and G. {Synnaeve} and A. {Joulin} and A. {Mohamed} and E. {Dupoux}},
  booktitle={ICASSP 2020 - 2020 IEEE International Conference on Acoustics, Speech and Signal Processing (ICASSP)}, 
  title={Libri-Light: A Benchmark for ASR with Limited or No Supervision}, 
  year={2020},
  pages={7669-7673},
  note = {\url{https://github.com/facebookresearch/libri-light}},
}

@article{chen2022wavlm,
  title={Wavlm: Large-scale self-supervised pre-training for full stack speech processing},
  author={Chen, Sanyuan and Wang, Chengyi and Chen, Zhengyang and Wu, Yu and Liu, Shujie and Chen, Zhuo and Li, Jinyu and Kanda, Naoyuki and Yoshioka, Takuya and Xiao, Xiong and others},
  journal={IEEE Journal of Selected Topics in Signal Processing},
  volume={16},
  number={6},
  pages={1505--1518},
  year={2022},
  publisher={IEEE}
}

@article{zeghidour2021soundstream,
  title={Soundstream: An end-to-end neural audio codec},
  author={Zeghidour, Neil and Luebs, Alejandro and Omran, Ahmed and Skoglund, Jan and Tagliasacchi, Marco},
  journal={IEEE/ACM Transactions on Audio, Speech, and Language Processing},
  volume={30},
  pages={495--507},
  year={2021},
  publisher={IEEE}
}

@inproceedings{zheng2025rethinking,
  title={Rethinking discrete tokens: Treating them as conditions for continuous autoregressive image synthesis},
  author={Zheng, Peng and Wang, Junke and Chang, Yi and Yu, Yizhou and Ma, Rui and Wu, Zuxuan},
  booktitle={Proceedings of the IEEE/CVF International Conference on Computer Vision},
  pages={17390--17400},
  year={2025}
}

\end{document}